\documentclass[preprint,aps,12pt,notitlepage,nofootinbib,tightenlines]{revtex4}
\usepackage[utf8]{inputenc}
\usepackage{amsmath}
\usepackage{bm}
\usepackage{times}
\usepackage{braket}
\usepackage{color}
\usepackage{epsfig}
\usepackage{slashed}
\usepackage{hyperref}
\usepackage{multirow}
\usepackage{booktabs}
\usepackage{array}
\usepackage{float}
\newcommand{\beq}{\begin{eqnarray}}
\newcommand{\eeq}{\end{eqnarray}}
\newcommand{\be}{\begin{equation}\begin{aligned}}
\newcommand{\ee}{\end{aligned}\end{equation}}

\newcommand{\gev}{\text{GeV}}

\definecolor{Red}{rgb}{1.,0.,0.}

\definecolor{Blue}{rgb}{0.,0.,1.}

\definecolor{nicered}{rgb}{0.7,0.1,0.1}
\definecolor{nicegreen}{rgb}{0.1,0.5,0.1}
\def\lsim{ {\ \lower-1.2pt\vbox{\hbox{\rlap{$<$}\lower6pt\vbox{\hbox{$\sim$}}}}\ } }
\def\gsim{ {\ \lower-1.2pt\vbox{\hbox{\rlap{$>$}\lower6pt\vbox{\hbox{$\sim$}}}}\ } }

\hypersetup{colorlinks,citecolor=nicegreen,linkcolor=nicered}
\begin{document}
%%%%%%%%%%%%%%%%%%%%%%%%%%%%%%%%%%%%%%%%%%%%%
\title{Single production of vectorlike $B$ quarks at the CLIC}
\author{Jin-Zhong Han\footnote{E-mail: hanjinzhong@zknu.edu.cn}, Jing Yang, Shuai Xu, Han-Kui Wang}
\affiliation{School of Physics and
Telecommunications Engineering, Zhoukou Normal University, Zhoukou 466001, China}

\begin{abstract}
The vector-like quarks are predicted in many new physics scenarios beyond the Standard Model~(SM)  and could be seen potential signatures of new physics  at the TeV energy scale. In this work,
we study single production of exotic singlet and doublet vectorlike bottom quarks (VLQ-$B$) at future Compact Linear Collider~(CLIC) via the process $e^{+}e^{-}\to B\bar{b}$ with the decay channel $B\to bZ$ and two types of modes: $Z\to \ell^{+}\ell^{-}$ and $Z\to \nu\bar{\nu}$. We calculate the cross sections of signal and relevant SM backgrounds. After a fast simulation of the signal and background events, the exclusion limit at  95\% confidence level  and $5\sigma$ discovery prospects on the parameters (the coupling strength $\kappa_{B}$  and the VLQ-$B$ mass) have been, respectively, presented at the future CLIC with centre of mass energy $\sqrt{s}=3$ TeV and integrated luminosity of 5~ab$^{-1}$.
\end{abstract}

\maketitle
\newpage
\section{Introduction}
  In order to solve the gauge hierarchy problem, the vector-like quarks (VLQs) are predicted to regulate the Higgs boson mass-squared divergence~\cite{DeSimone:2012fs} in several extensions of the Standard Model (SM), such as little Higgs models~\cite{ArkaniHamed:2002qy}, extra dimensions~\cite{Agashe:2006wa}, composite Higgs models~\cite{Agashe:2004rs}, and other TeV Scale new physics~(NP) models~\cite{He:1999vp,Wang:2013jwa,He:2001fz,He:2014ora}.  The left- and right-handed components of these new VLQs are transformed in the same properties under the SM electroweak symmetry group~\cite{Aguilar-Saavedra:2013qpa}. The VLQs are therefore not excluded by present searches, unlike a fourth
generation of SM quarks that is ruled out by electroweak precision measurements~\cite{Kribs:2007nz,Banerjee:2013hxa}.
Based on the electric charges of $+2/3e$~($T$ quark), $-1/3e$~($B$ quark), $+5/3e$~($X$ quark) or $-4/3e$~($Y$ quark), the VLQs could be grouped in multiplets, such as electroweak singlet [$T$, $ B$], electroweak doublets [ $\left(X,T\right),\left(T,B\right)$ or $\left(B,Y\right)$], or electroweak triplets [$\left(X,T,B\right)$ or $\left(T,B,Y\right)$], and they could generate characteristic signatures at the current and future high-energy colliders, see e.g.~\cite{Atre:2011ae,Buchkremer:2013bha,Barducci:2017xtw,Cacciapaglia:2018qep,Fuks:2016ftf,Yang:2014usa,Liu:2018hum,Cacciapaglia:2018lld,
Aguilar-Saavedra:2019ghg,Wang:2020ips,Zhang:2017nsn,Han:2017cvu,Liu:2017rjw,Liu:2017sdg,Liu:2019jgp,Tian:2021oey,Moretti:2016gkr,Moretti:2017qby,
Carvalho:2018jkq,Roy:2020fqf,Buckley:2020wzk,Deandrea:2021vje,King:2021iah}.
Here we focus on the singlet or $(B, Y)$ doublet VLQ-$B$ quark, which only couples to third-generation SM quarks.

Using the Run 2 data, the direct searches for VLQ-$B$ have been performed and the constraints on VLQ-$B$ have been obtained at 95\% confidence level~(CL) by the ATLAS and CMS Collaborations~\cite{Aaboud:2018wxv,Aaboud:2018xpj,Aaboud:2018uek,Aaboud:2018ifs,Sirunyan:2018qau,Sirunyan:2019sza,Sirunyan:2018omb,Sirunyan:2020qvb,Aaboud:2018pii}. For instance, an analysis from CMS including single-lepton, dilepton, and multilepton final
states probed all decay modes of the VLQ-$B$, and excluded $B$ quark masses in the range $910-1240$ GeV~\cite{Sirunyan:2018omb}. Recently, the CMS Collaboration presented a search for VLQ-$B$ pair production in the fully hadronic final state~\cite{Sirunyan:2020qvb}, and excluded the $B$ masses up to 1570, 1390, and 1450 GeV for 100\% $B\to bh$, 100\% $B\to tZ$, and $BY$ doublet cases, respectively. The combination of searches utilizing various final states were performed by the ATLAS
Collaboration~\cite{Aaboud:2018pii},
and excluded values of the $B$ mass up to 1220, 1370, and 1140 GeV for the singlet, the $(T,B)$ doublet and $(B,Y)$ doublet cases, respectively.

Up to now, many phenomenological
analysis about the VLQ-$B$ have been performed at the LHC and LHeC~\cite{Nutter:2012an,Gong:2019zws,Gong:2020ouh}. Compared to the complicated QCD background at the
hadron colliders,  the future linear $e^{+}e^{-}$ collider has a particularly clear background environment, i.e., the final stage of Compact Linear Collider~(CLIC) are operating at an energy of 3 TeV~\cite{CLIC1,CLIC2,CLIC3,Dannheim:2013ypa}. Thus,
the high-energy linear collider is a precision machine that can accurately measure the characteristics of the new VLQs~\cite{Kitano:2002ss,Kong:2007uu,Senol:2011nm,Harigaya:2011yg,Guo:2014piv,Liu:2014pts,Qin:2021cxl,Han:2021kcr}.
In this work, we focus on the observability of the single VLQ-$B$ production at the CLIC via the process $e^{+}e^{-}\to B\bar{b}~(b\bar{B})$ combined with the $B \to bZ$ and the subsequent decay channels $Z\to \ell^{+}\ell^{-}$ and $Z\to \nu\bar{\nu}$, respectively. The advantage is that it has a higher potential than paired production due to less phase space suppression. In addition, this process can reveal the electroweak nature of the interaction between the VLQ-$B$ and $Z$ boson.  Therefore, we expect that once the VLQ-$B$ is discovered and its mass is determined, such work could serve as a complementary option for future high-energy linear colliders.

The paper is organized as follows. In section II, we give a brief description of the simplified model including the VLQ-$B$ with electrical charge $-1/3$, and discuss its single production at the CLIC. In section III we investigate the signal and discovery potential of the VLQ-$B$ in the $Zb$ decay channel at the CLIC. Finally, we conclude in section IV.

\section{Vector-like bottom quark in the simplified model}

The generic parametrization of an effective Lagrangian of VLQ-$B$ can be expressed as (showing only the couplings relevant for our analysis):
\begin{eqnarray}
  \label{eq:vlq}
      \mathcal{L} = &&
        \kappa_B\bigg \{\sqrt{\frac{\zeta_i\xi^{B}_{W}}{\Gamma^0_{W}}} \frac{g}{\sqrt{2}} [\bar{B}_{L/R} W^+_\mu \gamma^\mu u^{\,i}_{L/R}]
      +  \sqrt{\frac{\zeta_i\xi^{B}_{Z}}{\Gamma^0_{Z}}} \frac{g}{2c_W} [\bar{B}_{L/R} Z_\mu \gamma^\mu d^i_{L/R}]   \nonumber \\
      && -  \sqrt{\frac{\zeta_i\xi^{B}_{H}}{\Gamma^0_{H}}} \frac{M}{v} [\bar{B}_{R/L} H d^i_{L/R}]
          \bigg \} + \text{h.c.} ,
\end{eqnarray}
where $g$ is the $SU(2)_L$ gauge coupling constant, $c_W$ is the usual cosine of the weak mixing
angle, $v\simeq 246$ GeV and $\xi^V$ parameters controlling the relative strengths of the $V$
couplings to top partners, and $\zeta_i$ parameters governing the mix of SM quark
generations $i$ in each coupling.
 $\xi$ and $\zeta$ are defined as $\sum_{V} \xi^{V} = 1$ and $\sum_{i} \zeta_i = 1$, meaning
$\zeta_i\xi^{V} = \textrm{BR}(B \rightarrow Vq_i)$~(for a detailed review, see~\cite{Buchkremer:2013bha}).

Although couplings of VLQs to first- and second-generation
SM quarks are not excluded~\cite{Atre:2008iu,Atre:2011ae}, much of the experimental and theoretical attention is on VLQs that
couple to third-generation SM quarks, as it is this generation that requires fine-tuning in the SM. Here we
focus on VLQ-$B$ that couples exclusively to third-generation SM quarks. Our results are interpreted assuming that
the B quark belongs to a singlet or doublet representation and that it decays exclusively
to SM particles.  In this case, the
singlet $B$ quark has three different decay channels into SM particles: $tW$, $bZ$ and $bH$. Using the equivalence theorem~\cite{ET-hjh,He:1992nga,He:1993yd,He:1994br,He:1996rb,He:1996cm} the branching fractions for these three decay modes are 0.5, 0.25 and 0.25, respectively. The $B$ doublet can decay to $bZ$ or $bH$, each with a branching
fraction of 0.5. Thus there are only two free parameters: the $B$ quark mass $m_B$ and the coupling strength $\kappa_{B}$.

\section{Event generation and discovery potentiality}
\begin{figure}[h]
\centering
\includegraphics[width = 12cm ]{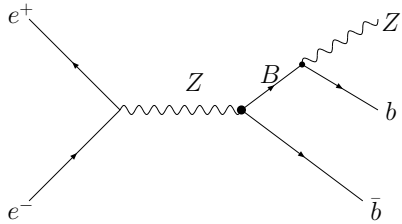}
\vspace{-12cm}
\caption{Representative Feynman diagrams of the process $e^{+}e^{-}\to B(\to Zb)\bar{b}$.}
\label{fig:fey}
\end{figure}

In Fig.~\ref{fig:fey}, we show the leading order Feynman diagram of the process $e^{+}e^{-}\to B\bar{b}$ with the decay mode $B\to Zb$.
%%%%%%%%%%%%%%%%%%%%%%%%%

In order to make a prediction for the signal, we calculate the cross section for the process $e^{+}e^{-}\to B\bar{b}~(b\bar{B})$ times the branching ratio of $B\to bZ$ at leading order~(LO) by using MadGraph5-aMC$@$NLO \cite{mg5}.  The numerical values of the input parameters are taken from~\cite{pdg}.
%%Fig.2 %%%%%%%%%%%%%%%%%%%%
\begin{figure}[h]
\begin{center}
\vspace{-0.5cm}
\centerline{\epsfxsize=11cm \epsffile{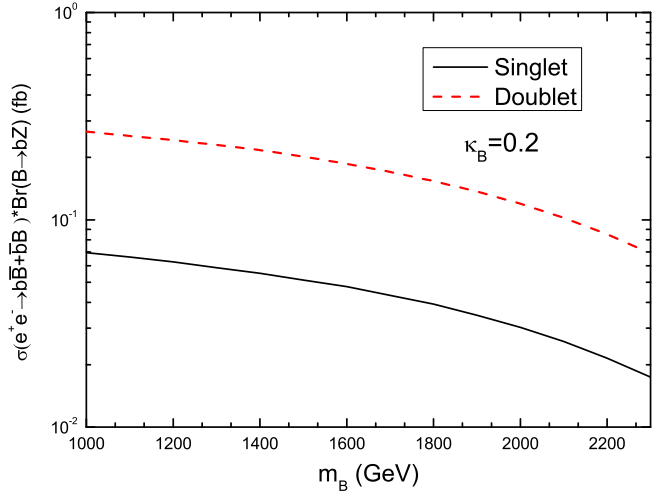}}
\caption{Total cross sections as a function of $m_B$ with $\kappa_{B}=0.2$ and two cases. }
\label{cs}
\end{center}
\end{figure}
%%%%%%%%%%%%%%%%%%%%%%%%%

In Fig.~\ref{cs}, we show the dependence of the cross sections $\sigma(e^{+}e^{-}\to B\bar{b}+b\bar{B})\times Br(B\to bZ)$ on the $B$ quark mass
$m_B$ at a 3 TeV CLIC for  $\kappa_{B}=0.2$. As the $B$ quark mass grows, the cross section of single production decreases slowly due to a larger phase space.
For $\kappa_{B}=0.2$ and $m_B=1.5~(2)$ TeV, the cross section can reach about 0.05~(0.3) fb for the singlet case and 0.2~(0.12) fb for the doublet case, respectively.
Obviously, the cross section of single $B-$quark production is proportional to the square of the coupling strength $\kappa_{B}$ for a given $B$ quark mass.

In next section, we will perform the Monte Carlo simulation and explore the discovery potentiality of VLQ-$B$
through the subsequent leptonic decay channel $Z\to \ell^{+}\ell^{-}$ and the invisible decay channel $Z\to \nu\bar{\nu}$, respectively.

Monte Carlo event simulations for signal and SM background are interfaced to Pythia 8.20~\cite{pythia8}  for fragmentation and showering. All event samples are fed into the Delphes 3.4.2~\cite{deFavereau:2013fsa} with the CLIC detector card designed for 3 TeV~\cite{Leogrande:2019qbe}. In our analysis, jets are clustered with the Valencia Linear Collider~(VLC) algorithm~\cite{Boronat:2014hva,Boronat:2016tgd} in exclusive mode with a fixed number of jet~($N=2$ where $N$ corresponds to the number of partons expected in the final state) and fixed one size parameter $R=0.7$. The $b$-tagging efficiency is taken as the loose working points with 90\% b-tagging efficiency in order not to excessively reduce the signal efficiency. The misidentification rates are given as a function of energy and pseudorapidity, i.e., in a bit where $E>500$ GeV and $1.53<|\eta|\leq 2.09$, misidentification rates are $5\times 10^{-2}$. Finally, event analysis is performed by using MadAnalysis5~\cite{ma5}.

\subsection{The decay channel $Z\to \ell^{+}\ell^{-}$}
In this subsection, we analyze the signal and background events at the 3 TeV CLIC through the $Z\to \ell^{+}\ell^{-}$~($\ell=e,\mu$) decay channel.
\be
e^{+}e^{-}\to B\bar{b}\to Zb\bar{b}\to \ell^{+}\ell^{-}b\bar{b}.
\ee
For this channel, the typical signal is two $b$-jet and two Opposite-Sign Same-Flavor~(OSSF) leptons. The dominant SM backgrounds come from the SM processes  $e^{+}e^{-}\to b\bar{b}\ell^{+}\ell^{-}$, and $e^{+}e^{-}\to jj\ell^{+}\ell^{-}$ with the cross sections of 2.74~fb, and 8.39~fb, respectively. Note that the contribution from the processes $e^{+}e^{-}\to H(\to b\bar{b})Z$, $e^{+}e^{-}\to Z(\to b\bar{b})Z$, $e^{+}e^{-}\to b\bar{b}Z$ and $e^{+}e^{-}\to q\bar{q}Z$ are also included with the decay mode $Z\to \ell^{+}\ell^{-}$.

To identify objects, we choose the basic cuts at parton level for the signals and SM backgrounds as follows:
 \be
p_{T}^{\ell}>~20~\gev,\quad
  p_{T}^{j/b}>~30~\gev,\quad
 |\eta_{\ell/b/j}|<~2\\
  \ee
where $p_{T}^{\ell, b, j}$ are the transverse momentum of leptons, $b$-jets, and light jets, respectively.
 %%%%%%%%%%%%%%%%%%%%%%%%%%%%%%%%%%%%%%%%%%%%%%%%%%
\begin{figure*}[htb]
\begin{center}
\centerline{\hspace{2.0cm}\epsfxsize=9cm\epsffile{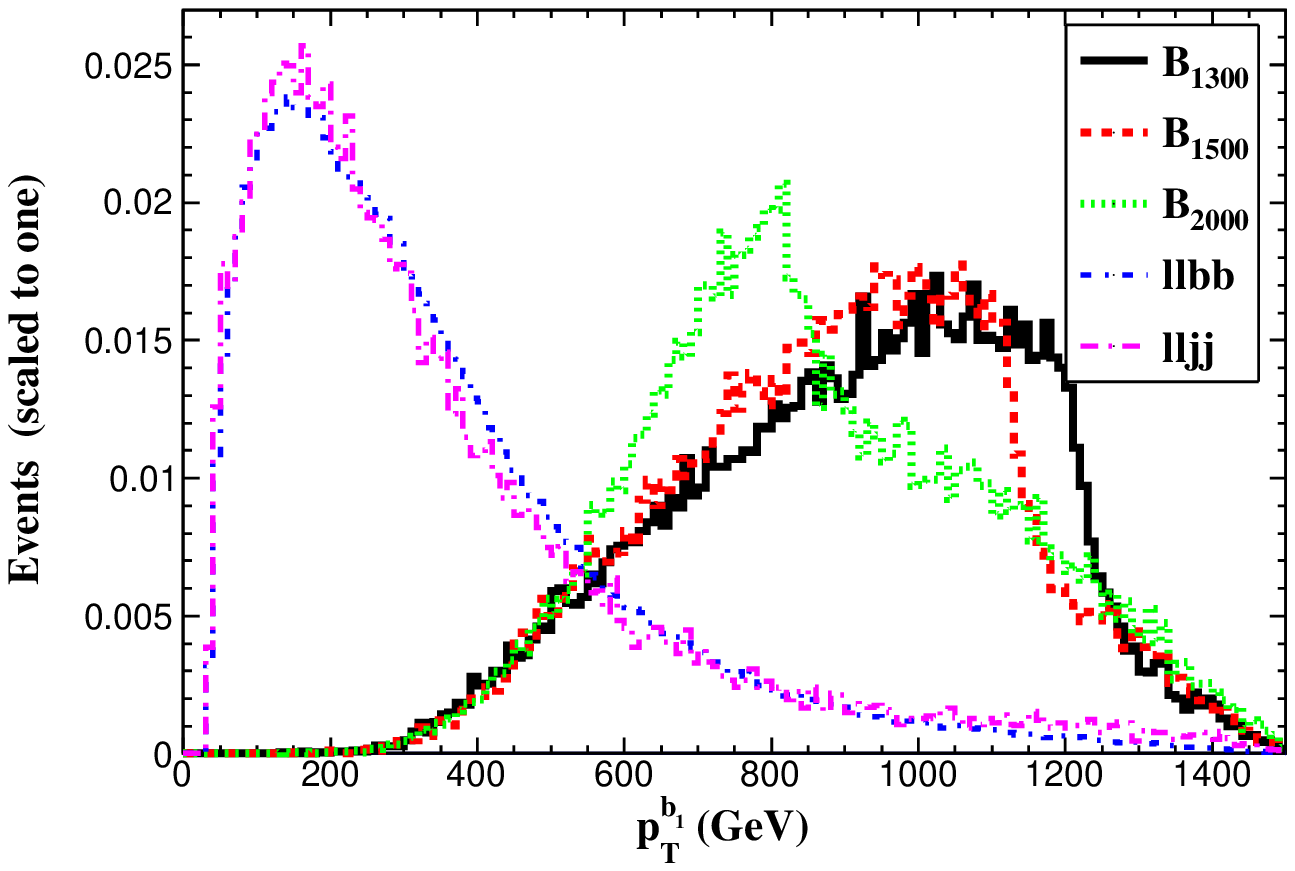}
\hspace{-2.0cm}\epsfxsize=9cm\epsffile{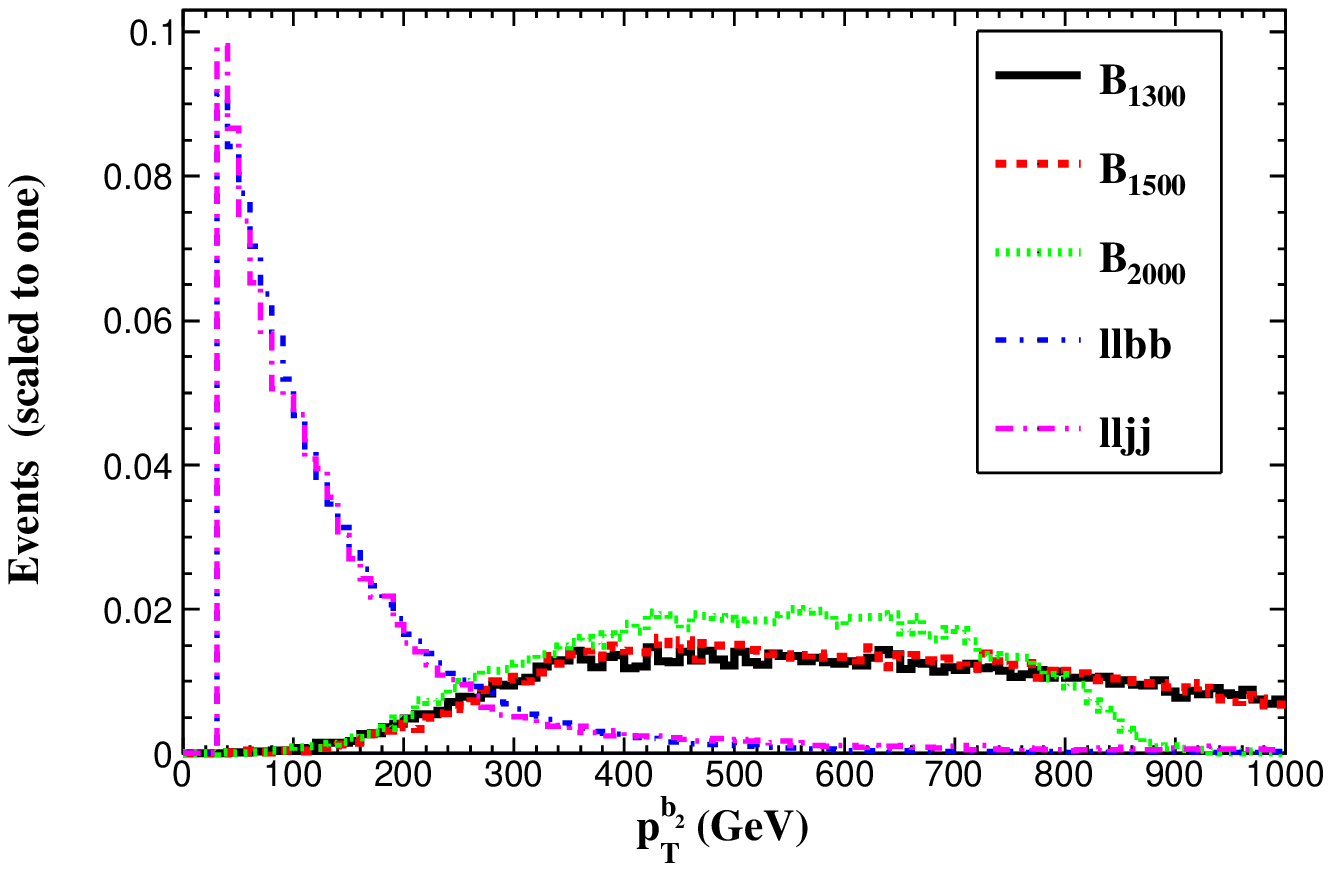}}
\centerline{\hspace{2.0cm}\epsfxsize=9cm\epsffile{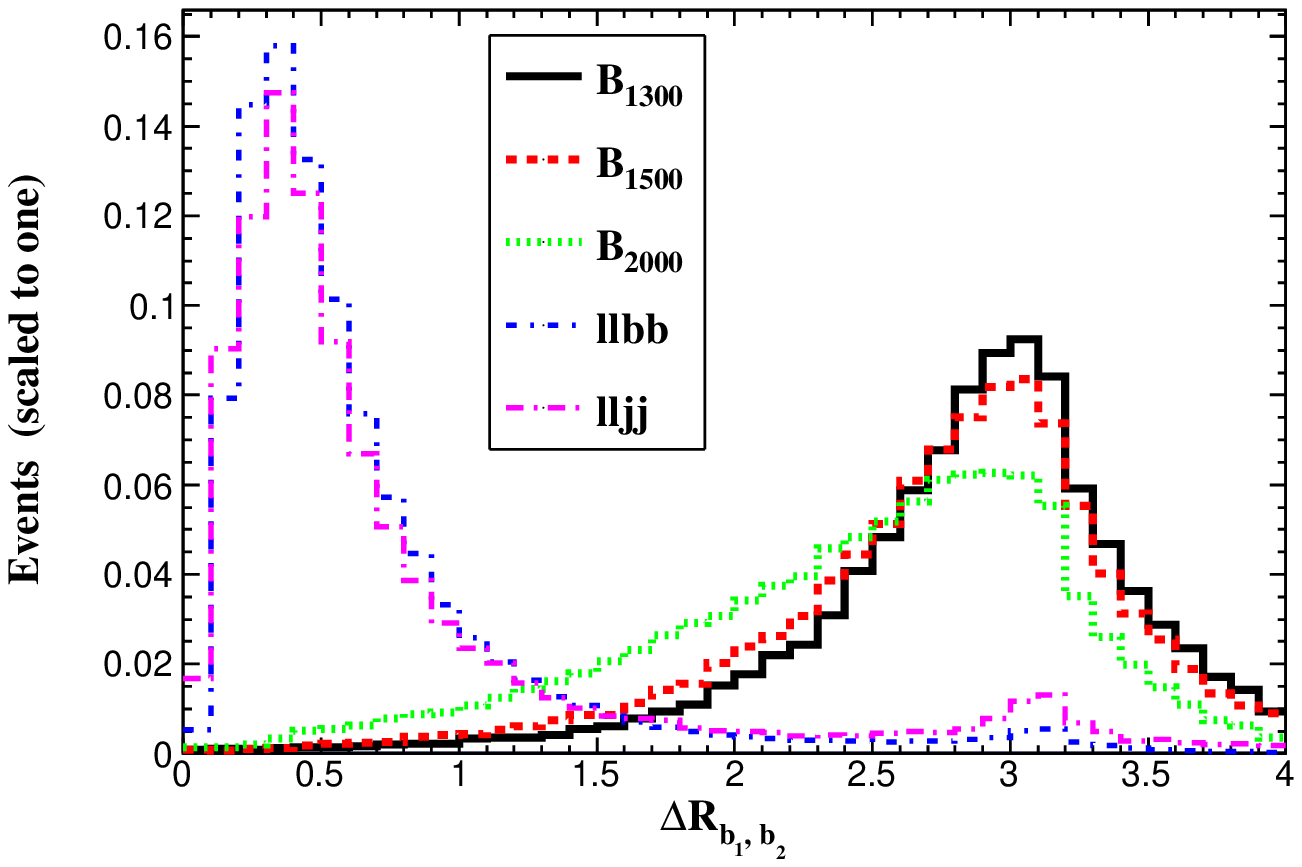}
\hspace{-2.0cm}\epsfxsize=9cm\epsffile{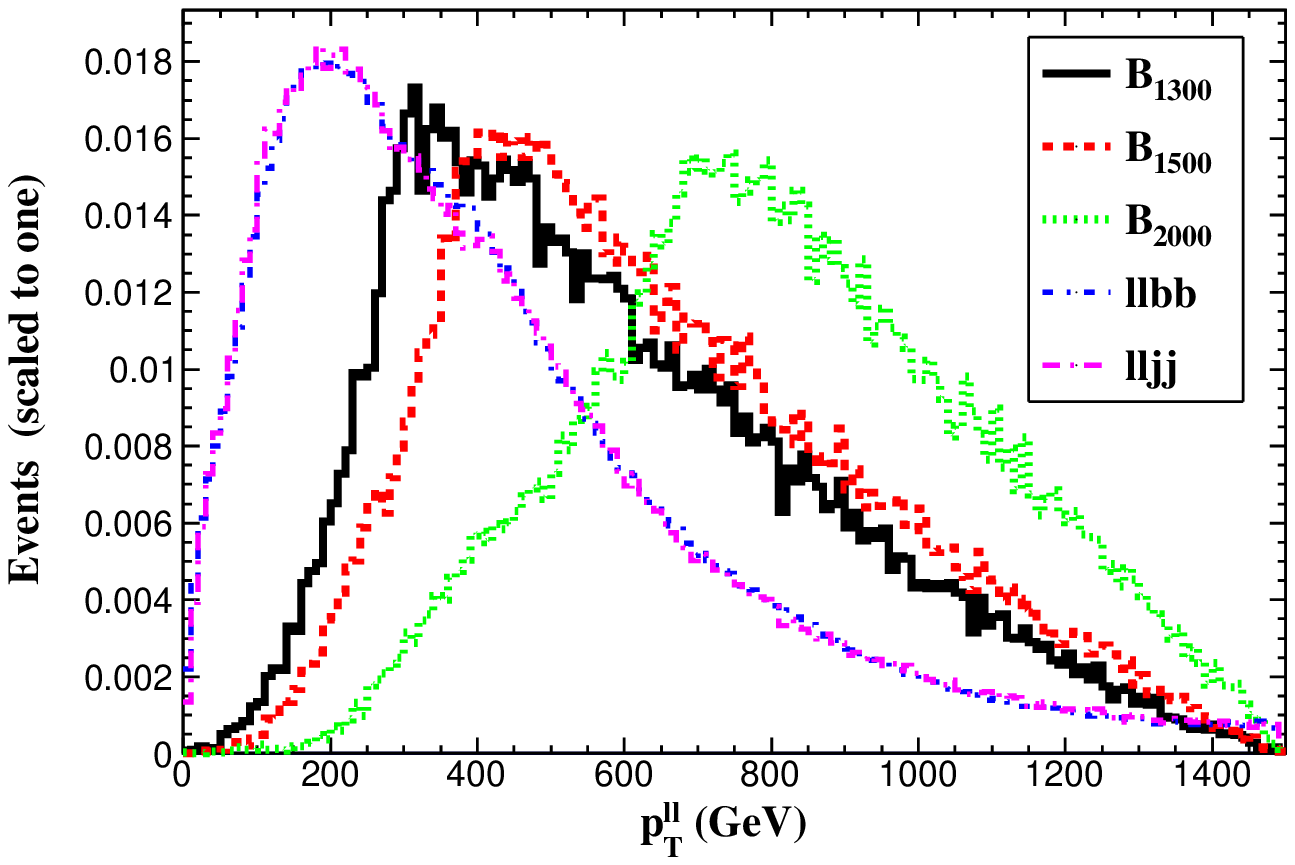}}
\centerline{\hspace{2.0cm}\epsfxsize=9cm\epsffile{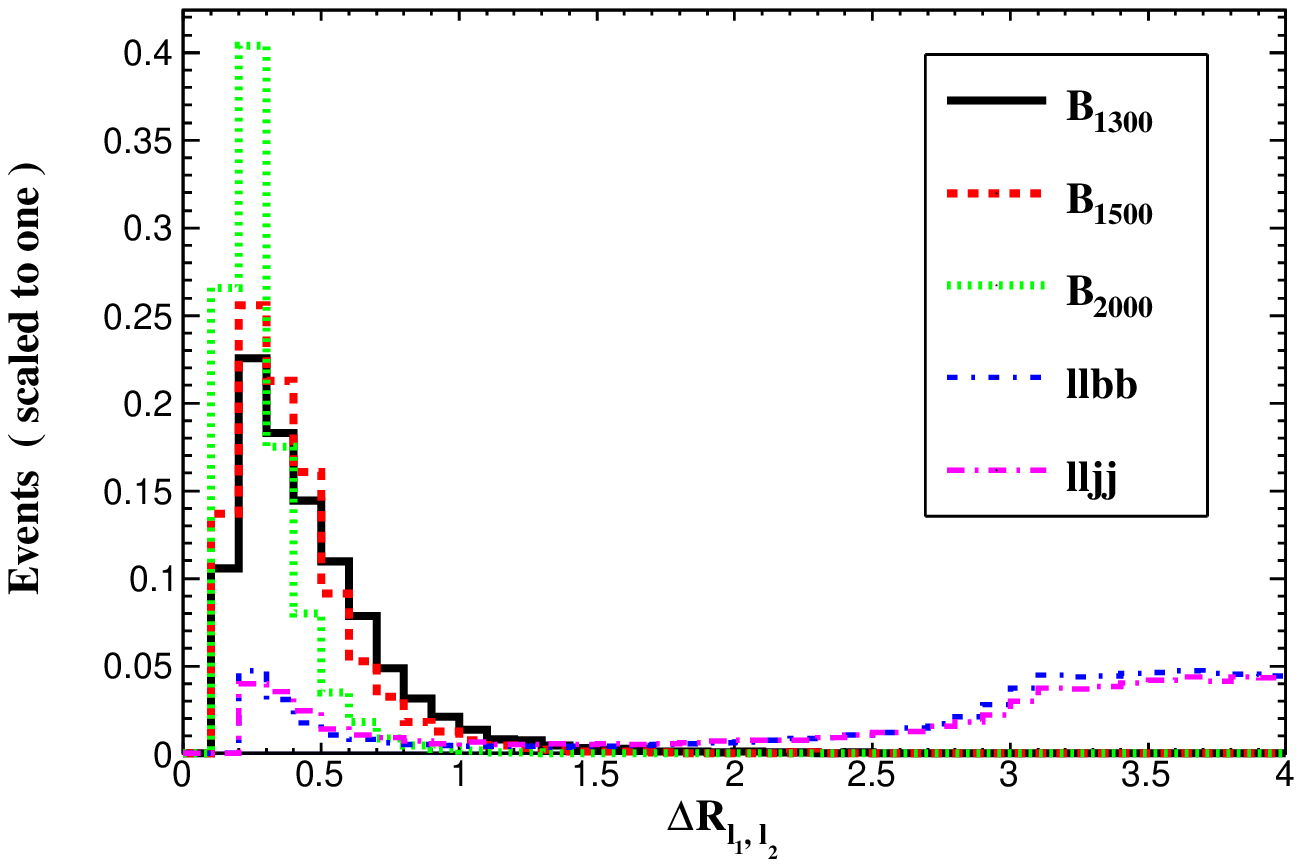}
\hspace{-2.0cm}\epsfxsize=9cm\epsffile{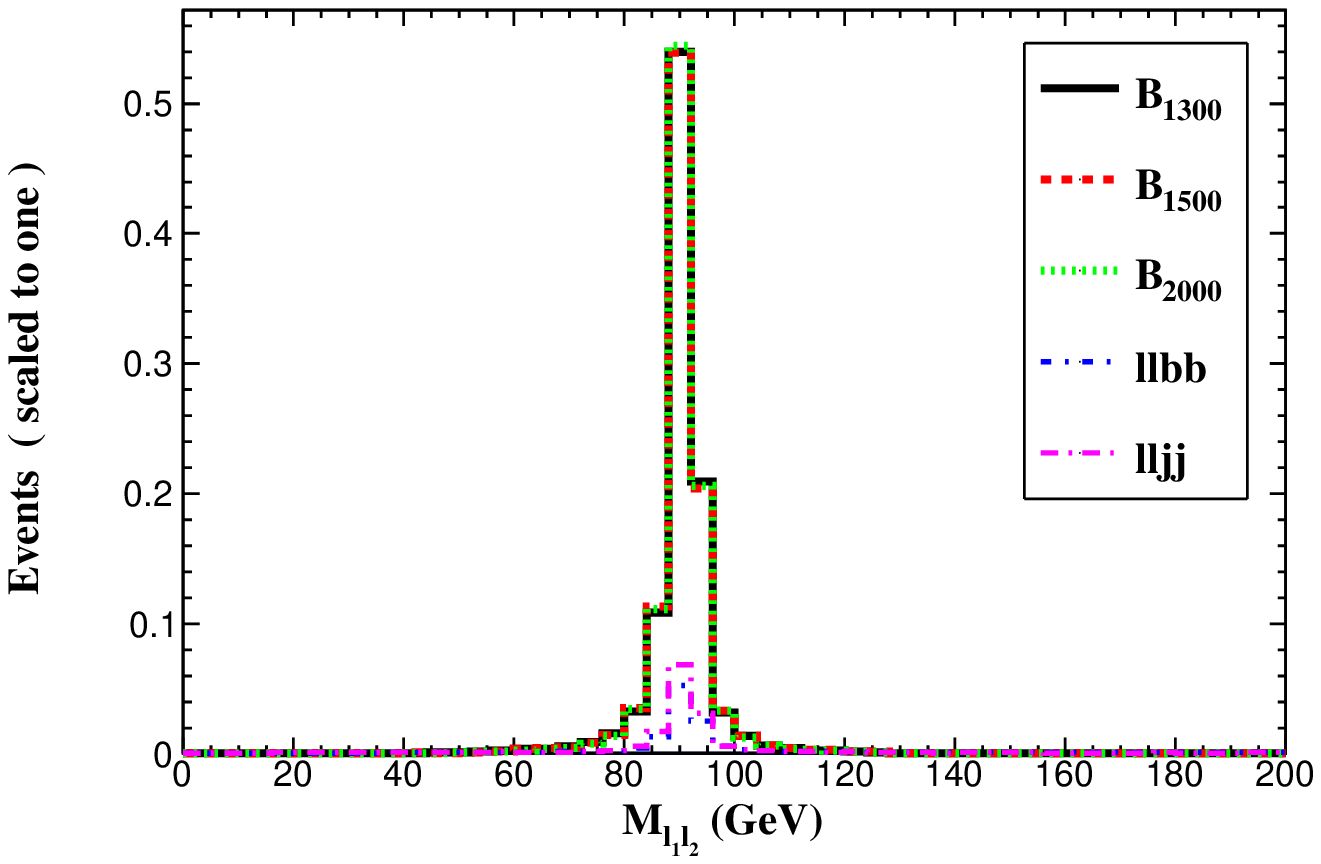}}
\caption{Normalized distributions for the signals (with $m_{B}$=1300, 1500 and 2000 GeV) and SM backgrounds at the CLIC. }
\label{distribution1}
\end{center}
\end{figure*}

For the signal, the leptons $\ell_{1}$ and $\ell_{2}$ are two OSSF leptons that are assumed to be the product of the $Z$-boson decay, and at least two $b$-tagged jet are present.
In Fig.~\ref{distribution1}, we plot some differential distributions  for  signals and SM backgrounds at the CLIC, such as the transverse momentum distributions of the leading and subleading $b$-jets~($p_{T}^{b_{1}}, p_{T}^{b_{2}}$), the separations $\Delta R_{b_{1},b_{2}}$, the transverse momentum distributions of the leading and subleading leptons~($p_{T}^{\ell_{1}\ell_{2}}$), the separations $\Delta R_{\ell_{1},\ell_{2}}$, and the invariant mass distribution for the $Z$ boson $M_{\ell_{1}\ell_{2}}$. Due to the larger mass of VLQ-$B$, the decay products of VLQ-$B$ are highly boosted. Therefore,
the $p_{T}^{l,b}$ peaks of the signals are larger than those of the SM backgrounds, and the lepton pairs of the
signal are much closer.
Based on these kinematical distributions, we can impose
 the following set of cuts:
 \begin{itemize}
\item Cut-1: There are exactly two isolated leptons and two $b$-tagged jets.
\item Cut-2: The transverse momenta of the leading and sub-leading $b$-jet are required $p_{T}^{b_{1}}> 500 \rm ~GeV$ and $p_{T}^{b_{2}}> 250 \rm ~GeV$ with $\Delta R_{b_{1},b_{2}}> 1.2$.
\item Cut-3: The transverse momenta of the leading and sub-leading leptons are required $p_{T}^{l_{1}l_{2}}> 300 \rm ~GeV$, the invariant mass of the $Z$ boson is required to have $|M_{\ell_{1}\ell_{2}}-m_{Z}|< 10 \rm ~GeV$ with $\Delta R_{\ell_{1},\ell_{2}}< 1$.
\end{itemize}

\begin{table}[ht!]
\fontsize{12pt}{8pt}\selectfont \caption{Cut flow of the cross sections (in $10^{-3}$~fb) for the signals and SM backgrounds at the CLIC with $\kappa_{B}=0.2$ and three typical $B$ quark masses for the singlet case and doublet case~(in the bracket). \label{cutflow1}}
\begin{center}
\newcolumntype{C}[1]{>{\centering\let\newline\\\arraybackslash\hspace{0pt}}m{#1}}
{\renewcommand{\arraystretch}{1.5}
\begin{tabular}{C{2.0cm}| C{2.0cm} |C{2.0cm} |C{2.0cm}|C{1.6cm}C{1.6cm}C{1.6cm} }
\hline
 \multirow{2}{*}{Cuts}& \multicolumn{3}{c|}{Signals}&\multicolumn{3}{c}{Backgrounds} \\ \cline{2-7}
&1300 GeV &1500 GeV& 2000 GeV  & $\ell^{+}\ell^{-}b\bar{b}$ &$\ell^{+}\ell^{-}jj$& Total\\   \cline{1-7} \hline
Basic&3.4~(15.2)&3.3~(13.2)&1.9~(7.6)&1869&1179&3048\\
Cut 1&2.37~(9.5)&2.1~(8.4)&1.2~(4.8)&689&103&792\\
Cut 2 &2.25~(9.0)&1.95~(7.8)&1.1~(4.4)&18.2&7.7&25.9\\
Cut3 &1.62~(6.5)&1.58~(6.3)&0.95~(3.8)&5.34&2.02&7.36
\\
\hline
\end{tabular} }
 \end{center}
 \end{table}

We present the cross sections of three typical signal ($m_B=1300, 1500, 2000$ GeV) and the relevant
backgrounds after imposing
the cuts in Table~\ref{cutflow1}.
One can see that all the SM backgrounds are suppressed very efficiently with the cross section of about 0.01~fb, while the signals still have a relatively good efficiency at the end of the cut flow. The total SM  background comes from the $e^{+}e^{-}\to b\bar{b}\ell^{+}\ell^{-}$ process, with a total cross section of $5.34\times 10^{-3}$~fb.

\subsection{The decay channel $Z\to \nu\bar{\nu}$ }
In this subsection, we analyze the signal and background events through the invisible decays $Z\to \nu\bar{\nu}$ decay channel.
\be
e^{+}e^{-}\to B\bar{b}\to Zb\bar{b}\to b\bar{b}+\slashed E_{T}.
\ee
For this channel, the main SM backgrounds come from the processes
$e^{+}e^{-}\to \nu\bar{\nu}b\bar{b}$ and $e^{+}e^{-}\to \nu\bar{\nu}jj$ with the cross sections of 0.26~fb and 0.61~fb, respectively. Note that the contribution from the processes $e^{+}e^{-}\to HZ$, $e^{+}e^{-}\to \nu_{e}\bar{\nu}_{e}H$, $e^{+}e^{-}\to \nu_{e}\bar{\nu}_{e}Z$, and $e^{+}e^{-}\to Zb\bar{b}$ are also included with the decay mode $Z\to \nu\bar{\nu}$ and $H\to b\bar{b}$.

In our simulations, we apply the following basic cuts on the signal and background events  at parton level:
\be
p_{T}^{b/j}>~30~\gev,\quad |\eta_{b/j}|<~2, \quad \slashed E_{T}> 100~\gev. \\
  \ee

 %%%%%%%%%%%%%%%%%%%%%%%%%%%%%%%%%%%%%%%%%%%%%%%%%%
\begin{figure}[htb]
\begin{center}
\centerline{\hspace{2.0cm}\epsfxsize=10cm\epsffile{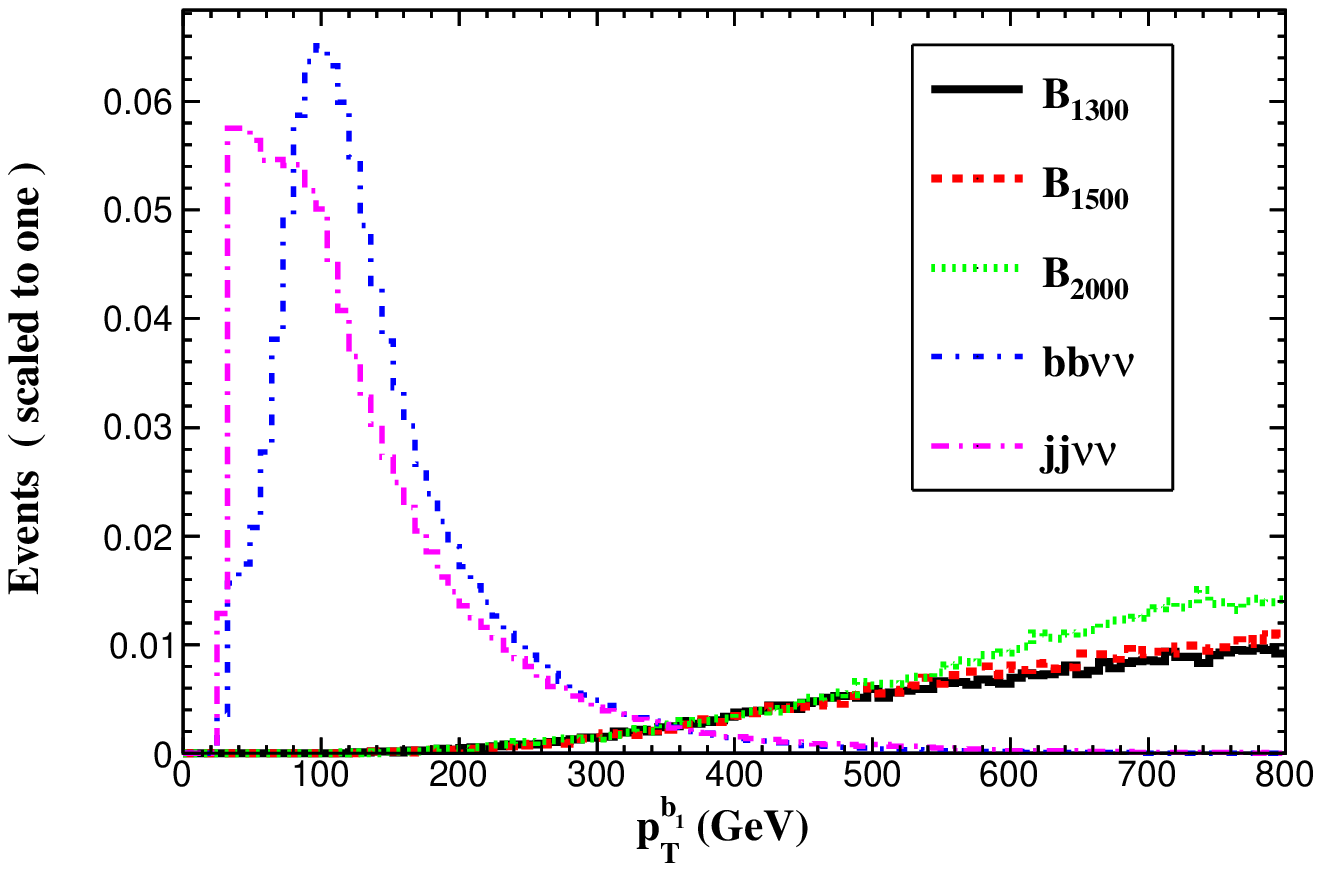}
\hspace{-2.0cm}\epsfxsize=10cm\epsffile{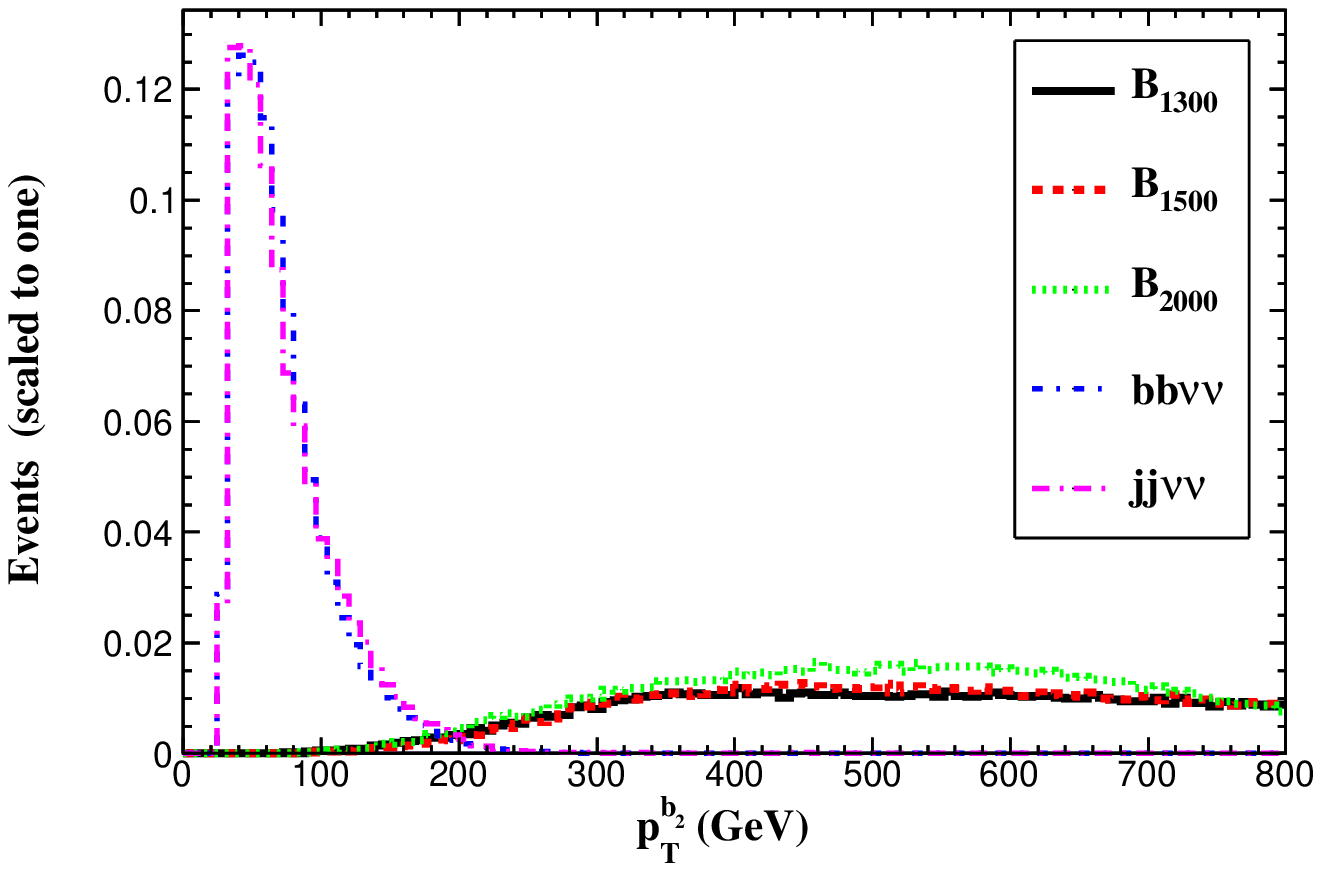}}
\centerline{\hspace{2.0cm}\epsfxsize=10cm\epsffile{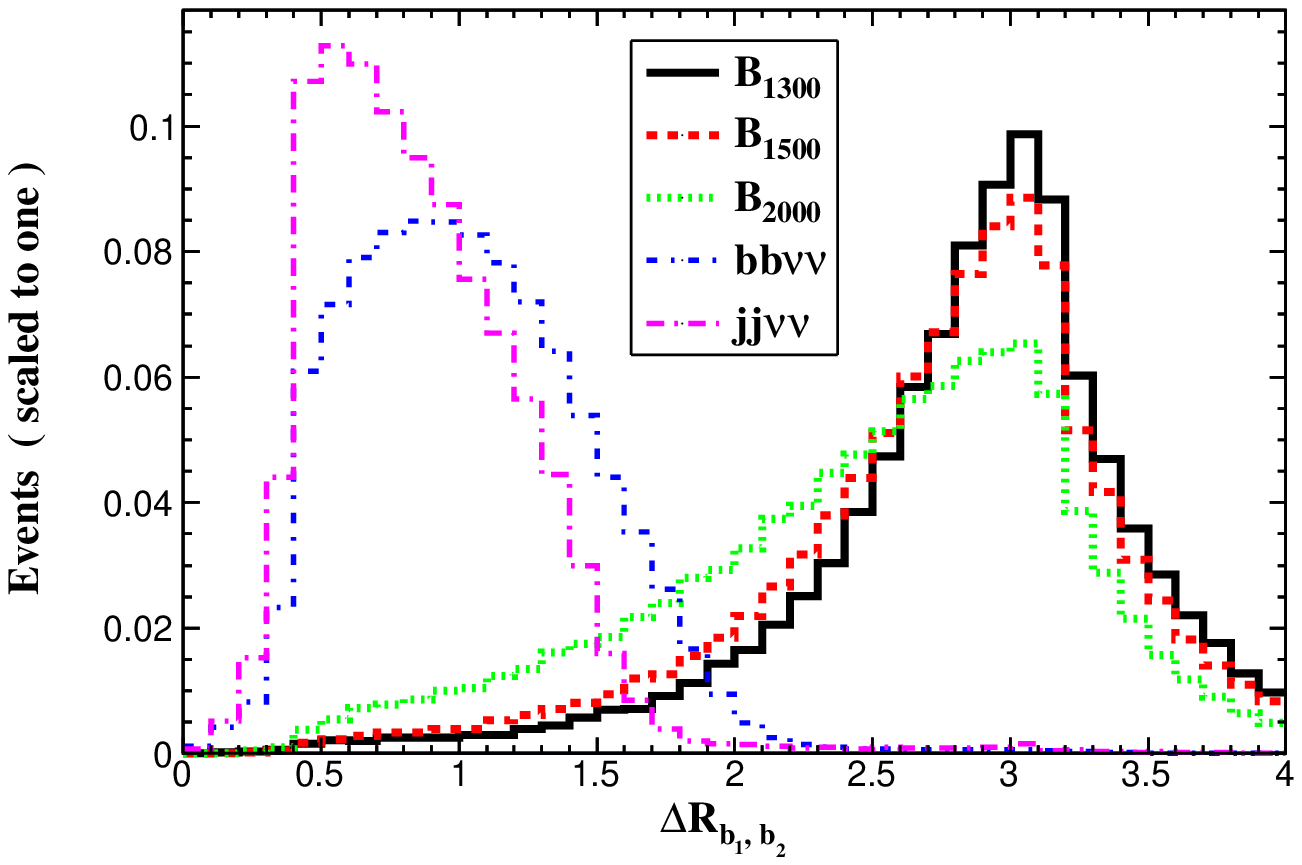}
\hspace{-2.0cm}\epsfxsize=10cm\epsffile{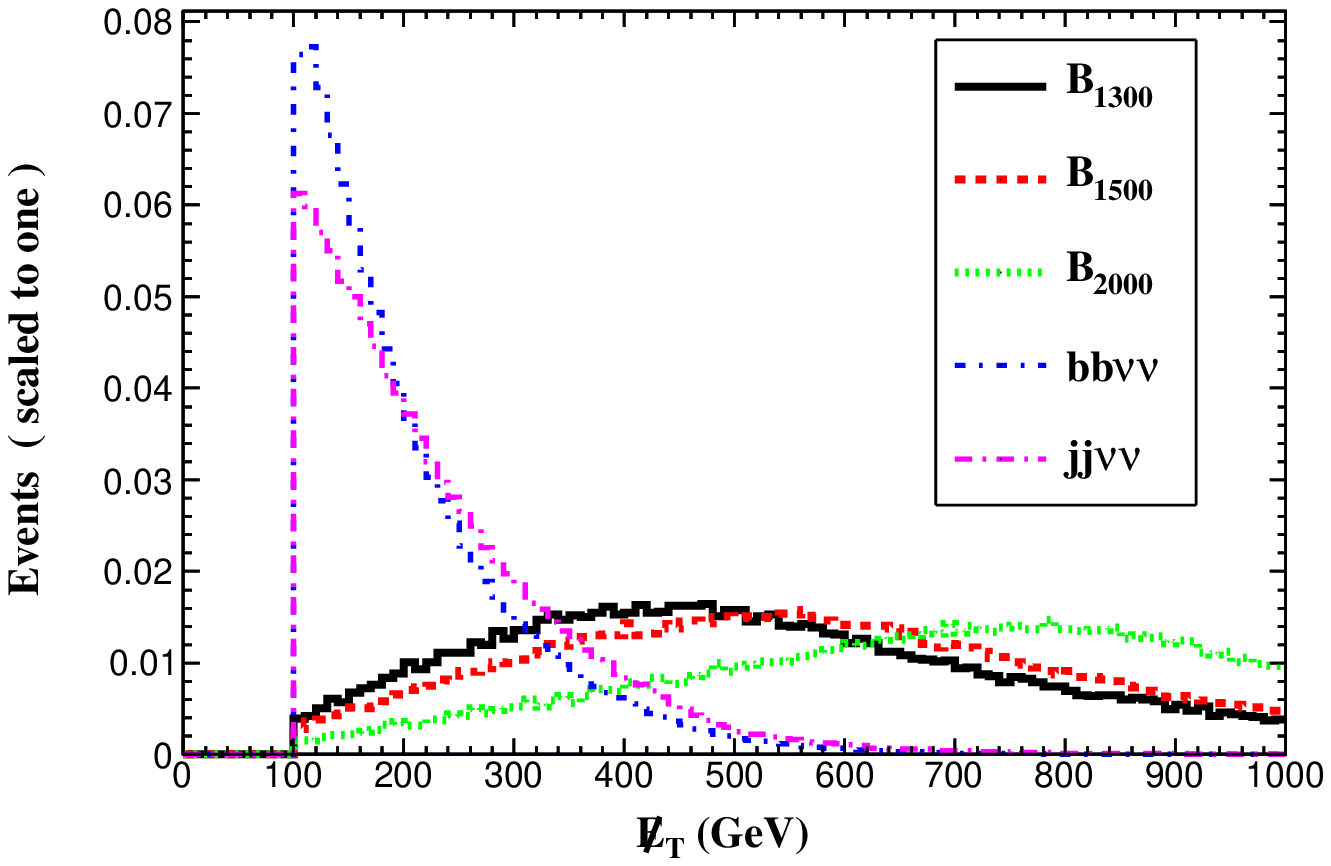}}
\caption{Normalized distributions for the signals and SM backgrounds for $Z\to \nu\bar{\nu}$ decay channel at the CLIC. }
\label{distribution2}
\end{center}
\end{figure}

Obviously, the signal events should contain large missing transverse energy $\slashed E_{T}$  from the boosted $Z$ boson.
In order to get some hints of further cuts for reducing the SM backgrounds, we analyzed the normalized distributions of  $p_{T}^{b_{1},b_{2}}$, $p_{T}^{b_{2}}$, $\Delta R_{b_{1},b_{2}}$, and $\slashed E_{T}$ for  signals and SM backgrounds as shown in Fig.~\ref{distribution2}.
Based on these kinematical distributions, a set of further cuts are given as:
 \begin{itemize}
\item Cut-1: There are at least two $b$-tagged jets and remove any electrons and muons.
\item Cut-2: The transverse momenta of the leading and sub-leading $b$-jet are required $p_{T}^{b_{1}}> 300 \rm ~GeV$ and $p_{T}^{b_{2}}> 200 \rm ~GeV$ with $\Delta R_{b_{1},b_{2}}> 1.5$.
\item
Cut-3: The transverse missing energy is required $\slashed E_{T}> 300 \rm ~GeV$.
\end{itemize}

\begin{table}[ht!]
\fontsize{12pt}{8pt}\selectfont \caption{Cut flow of the cross sections (in $10^{-3}$~fb) for the signals and SM backgrounds at the CLIC with $\kappa_{B}=0.2$ and three typical $B$ quark masses for the singlet case and doublet case~(in the bracket). \label{cutflow2}}
\begin{center}
\newcolumntype{C}[1]{>{\centering\let\newline\\\arraybackslash\hspace{0pt}}m{#1}}
{\renewcommand{\arraystretch}{1.5}
\begin{tabular}{C{2.0cm}| C{2.0cm} |C{2.0cm} |C{2.0cm}|C{1.8cm}C{1.8cm}C{1.8cm} }
\hline
 \multirow{2}{*}{Cuts}& \multicolumn{3}{c|}{Signals}&\multicolumn{3}{c}{Backgrounds} \\ \cline{2-7}
&1300 GeV &1500 GeV& 2000 GeV  & $\nu\bar{\nu}b\bar{b}$ &$\nu\bar{\nu}jj$&Total\\   \cline{1-7} \hline
Basic&11~(44)&~8.5~(38)&5.5~(22)&223590&50120&273710\\
Cut 1&8~(32)&7~(28)&4.1~(16.4)&128500&5359&133859\\
Cut 2 &7.5~(30)&6.5~(26)&3.6~(14.4)&206&25&231\\
Cut3 &6.1~(24.4)&5.6~(22.4)&3.3~(13.2)&92&12&104
\\
\hline
\end{tabular} }
 \end{center}
 \end{table}

We summarize the cross sections of three typical signal ($m_B=1300, 1500, 2000$ GeV) and the relevant
backgrounds after imposing
the cuts in Table~\ref{cutflow2}.
One can see that the total SM backgrounds are suppressed very efficiently, with a cross section of about 0.1~fb.

\subsection{Discovery and exclusion significance }

In order to see whether the signatures of VLQ-$B$ can be detected at the CLIC, we use the median significance  to estimate the expected discovery and exclusion significance~\cite{Cowan:2010js}:
\be
\mathcal{Z}_\text{disc} &=
  \sqrt{2\left[(s+b)\ln\left(\frac{(s+b)(1+\delta^2 b)}{b+\delta^2 b(s+b)}\right) -
  \frac{1}{\delta^2 }\ln\left(1+\delta^2\frac{s}{1+\delta^2 b}\right)\right]} \\
   \mathcal{Z}_\text{excl} &=\sqrt{2\left[s-b\ln\left(\frac{b+s+x}{2b}\right)
  - \frac{1}{\delta^2 }\ln\left(\frac{b-s+x}{2b}\right)\right] -
  \left(b+s-x\right)\left(1+\frac{1}{\delta^2 b}\right)},
 \ee
with
 \be
 x=\sqrt{(s+b)^2- 4 \delta^2 s b^2/(1+\delta^2 b)}.
  \ee
Here, the values of $s$ and $b$ were obtained by multiplying the total signal and SM background cross
sections, respectively,  by the integrated luminosity. $\delta$ is the percentage systematic error on the SM
background estimate.
In the limit of
$\delta \to 0$,  these expressions  can be simplified as
\be
 \mathcal{Z}_\text{disc} &= \sqrt{2[(s+b)\ln(1+s/b)-s]}, \\
 \mathcal{Z}_\text{excl} &= \sqrt{2[s-b\ln(1+s/b)]}.
\ee
In this work we choose two cases: no systematics ($\delta=0$) and a systematic
uncertainty of $\delta=10\%$.
%%% Fig.5 %%%%%%%%%%%%%%%%%%%%
\begin{figure}[htb]
\begin{center}
\vspace{-0.5cm}
\centerline{\epsfxsize=8cm \epsffile{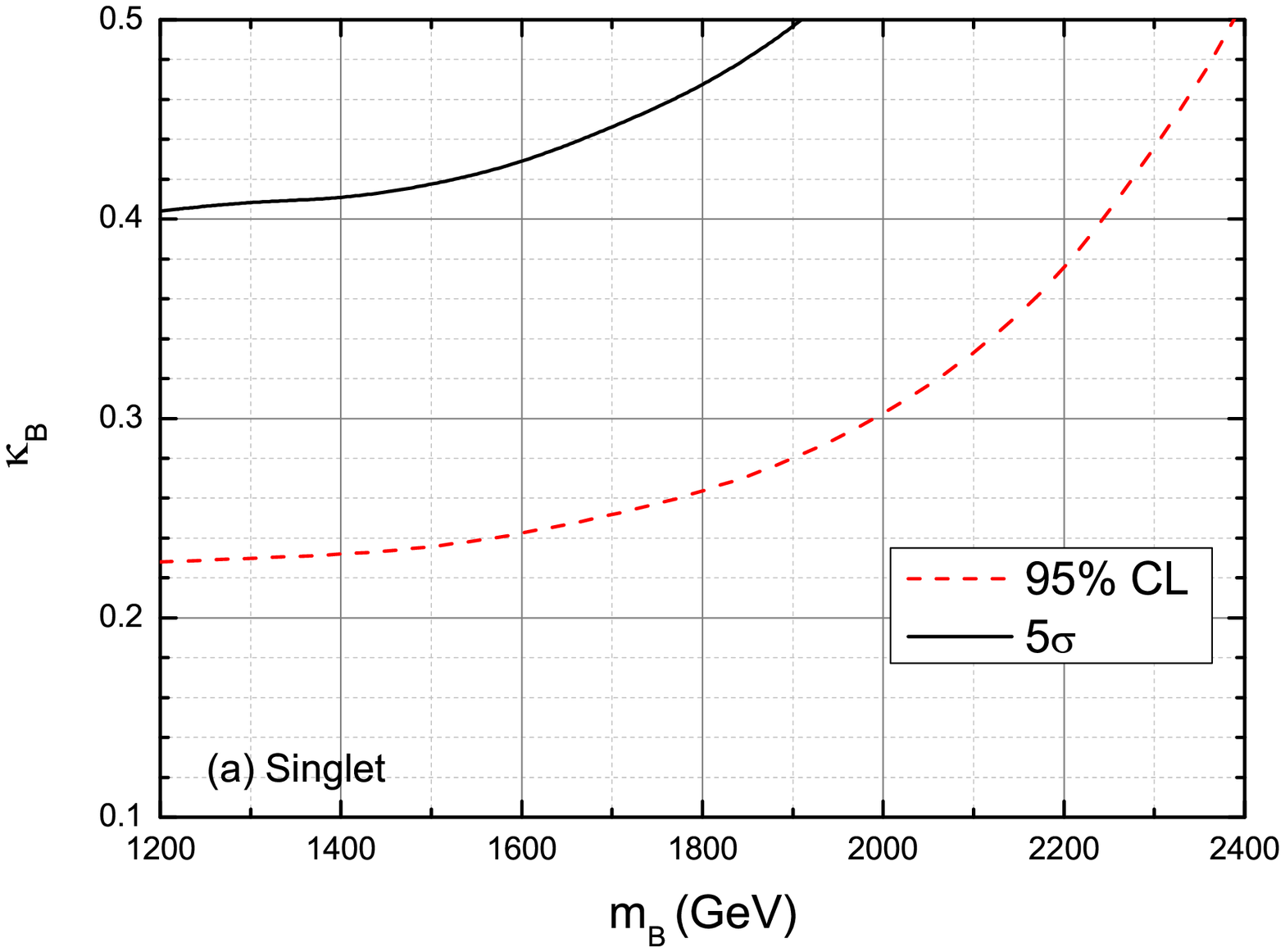}\epsfxsize=8cm \epsffile{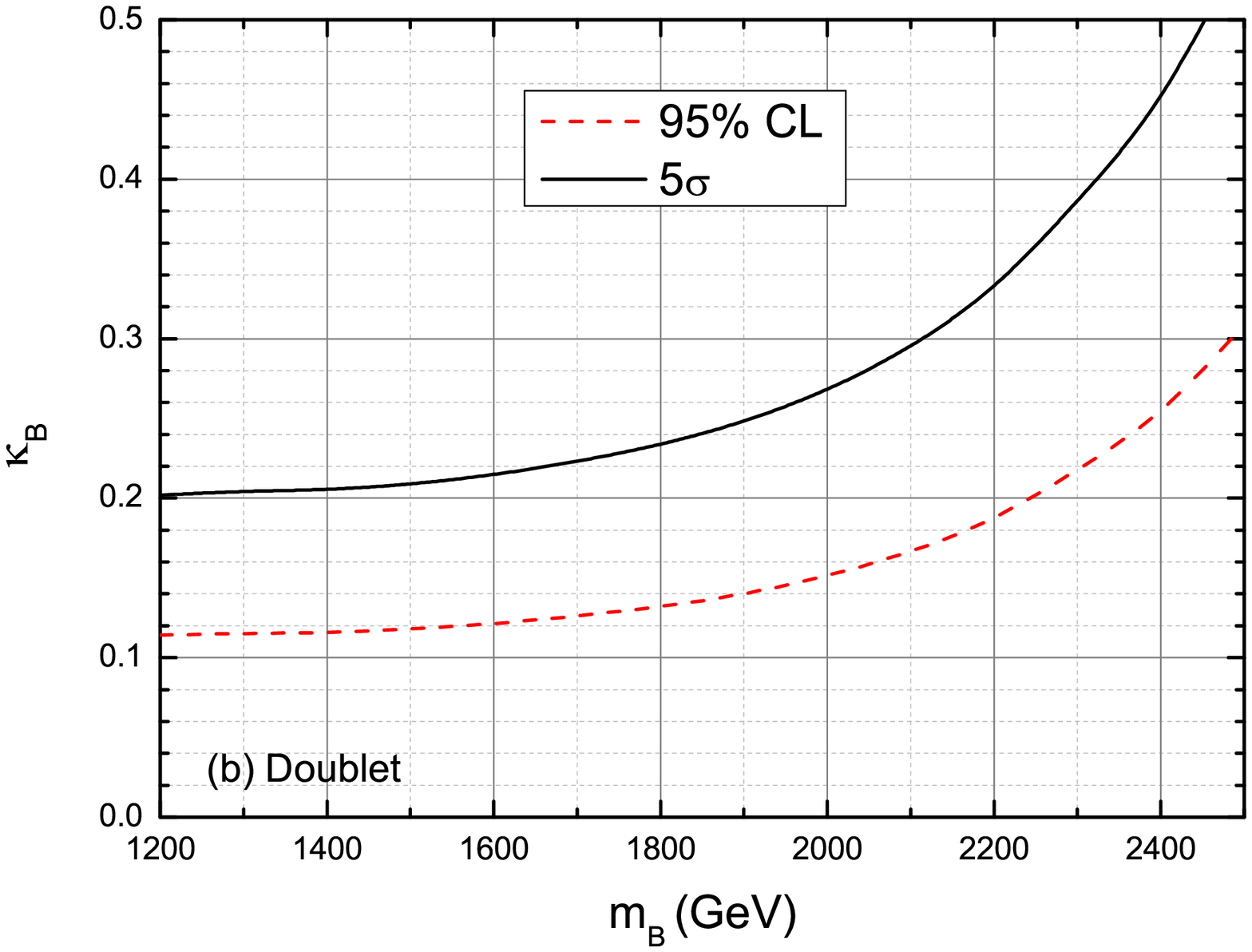}}
\caption{ The exclusion limit~(at $95\%$ CL) and discovery prospects  (at $5\sigma$) contour plots for the signal in $\kappa_{B}-m_{B}$ planes at 3 TeV CLIC with integral luminosity 5 ab$^{-1}$  for $Z\to \ell^{+}\ell^{-}$ decay channel. }
\label{fig5}
\end{center}
\end{figure}

%%% Fig.6 %%%%%%%%%%%%%%%%%%%%
\begin{figure}[htb]
\begin{center}
\vspace{-0.5cm}
\centerline{\epsfxsize=8cm \epsffile{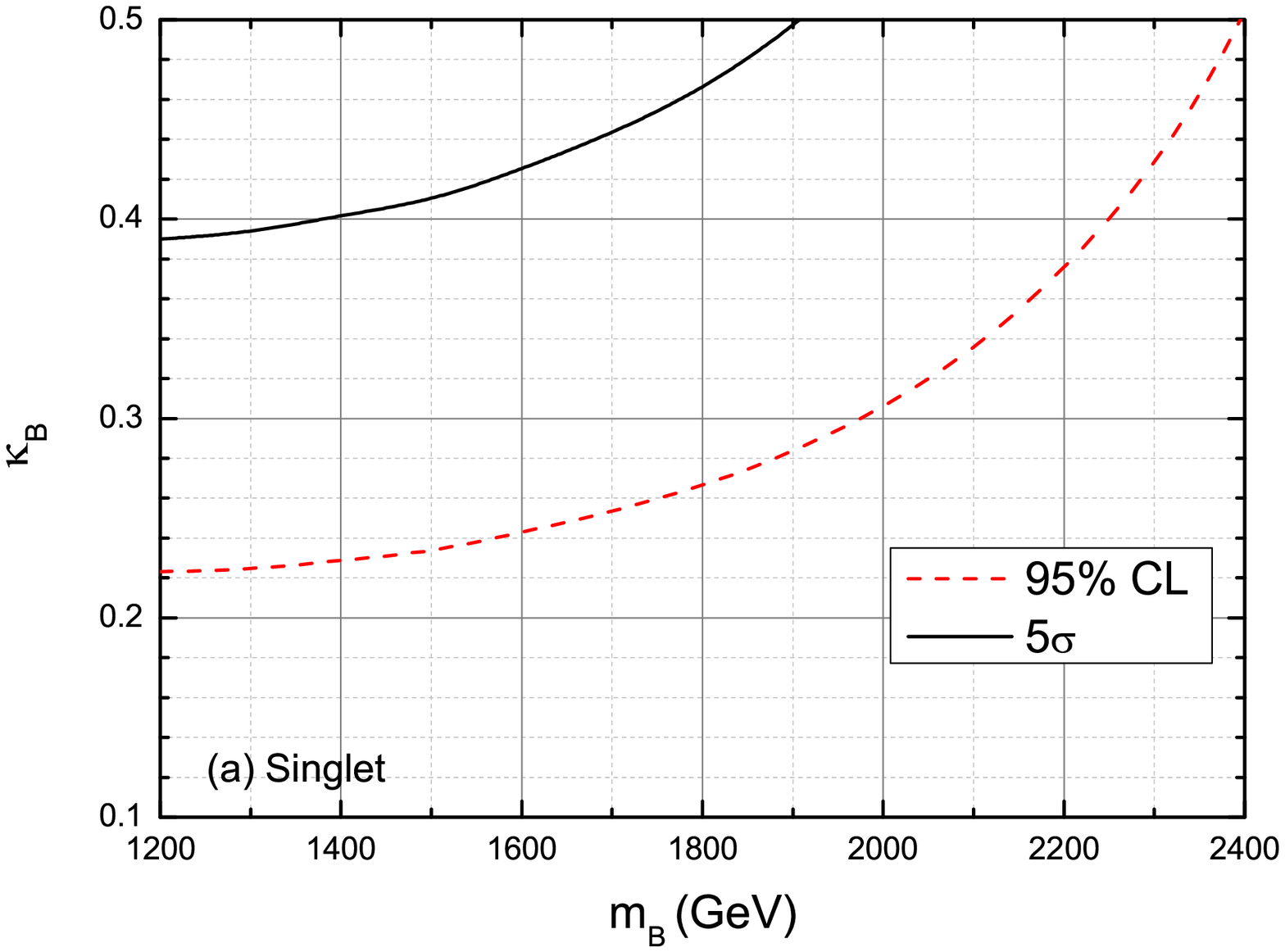}\epsfxsize=8cm \epsffile{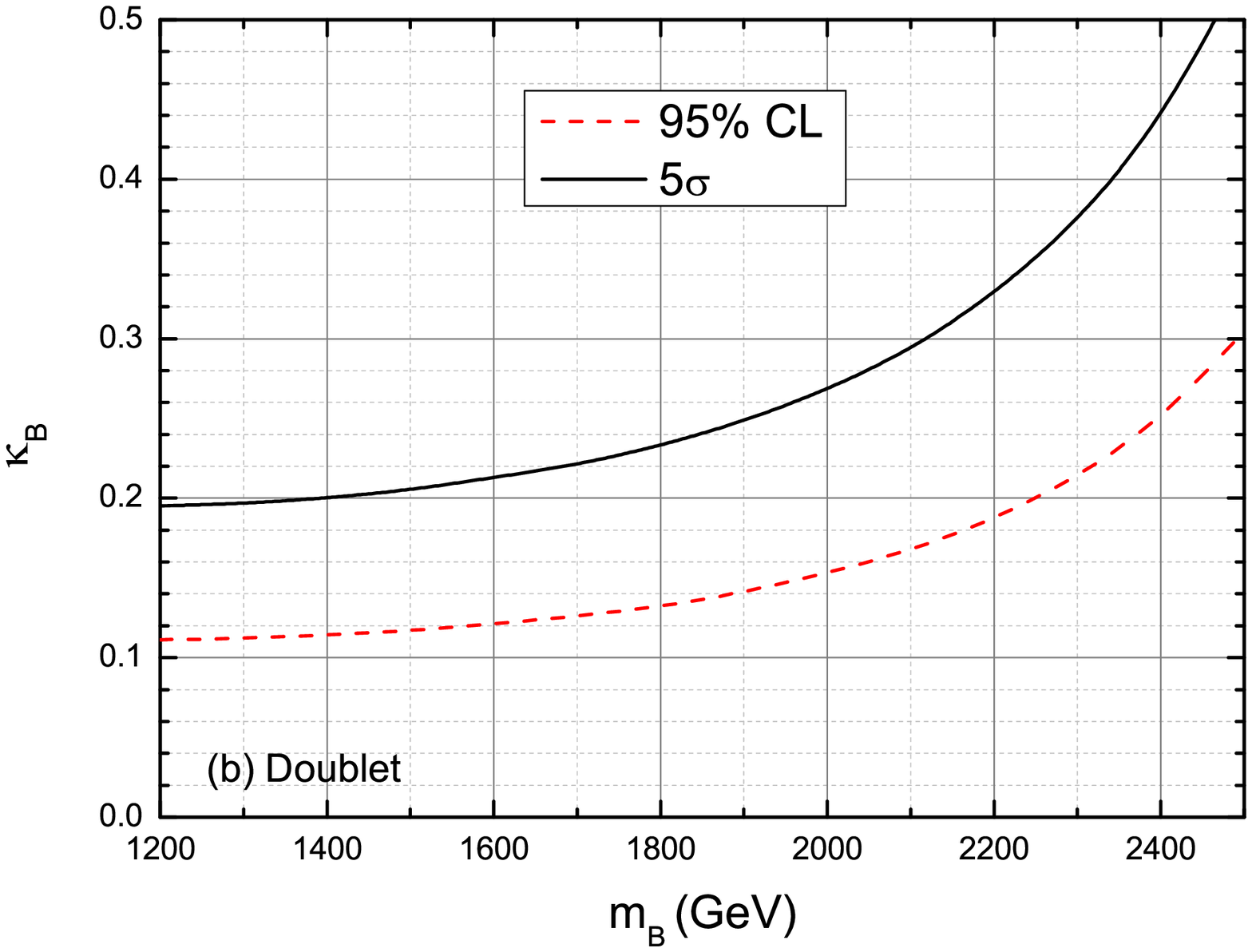}}
\caption{Same as Fig.~\ref{fig5} but for $Z\to \nu\bar{\nu}$ decay channel. }
\label{fig6}
\end{center}
\end{figure}

In Figs.~\ref{fig5}-\ref{fig6}, we plot the exclusion limit at 95\% confidence level~(CL) and $5\sigma$ sensitivity reaches for the coupling strength  $\kappa_{B}$ as a function of $m_B$ at 3~TeV CLIC with integral luminosity 5 ab$^{-1}$, respectively, for two decay channels without considering the effect of the systematic error. One finds that, for the $Z\to \ell^{+}\ell^{-}$ decay channel, the singlet~(doublet) VLQ-$B$ quarks can be  excluded in the region of $\kappa_{B}\in [0.23, 0.62]~([0.11, 0.31])$ and $m_B\in$ [1200 GeV, 2500 GeV]  at the 3 TeV CLIC with the integrated luminosity of 5 ab$^{-1}$, while the discover region can reach  $\kappa_{B}\in [0.4, 0.53]~([0.2, 0.55])$ and $m_B\in$ [1200 GeV, 2000 GeV]~([1200 GeV, 2500 GeV]).
Similarly, for $Z\to \nu\bar{\nu}$ decay channel, the singlet~(doublet) VLQ-$B$ quarks can be excluded in the region of $\kappa_{B}\in [0.22, 0.5]~([0.11, 0.3])$ and $m_B\in$ [1200 GeV, 2400 GeV]~([1200 GeV, 2500 GeV]),  the discover region can reach  $\kappa_{B}\in [0.39, 0.5]~([0.2, 0.44])$ and $m_B\in$ [1200 GeV, 1900 GeV]~([1200 GeV, 2400 GeV]).

%%% Fig.7 %%%%%%%%%%%%%%%%%%%%
\begin{figure}[htb]
\begin{center}
\vspace{-0.5cm}
\centerline{\epsfxsize=8cm \epsffile{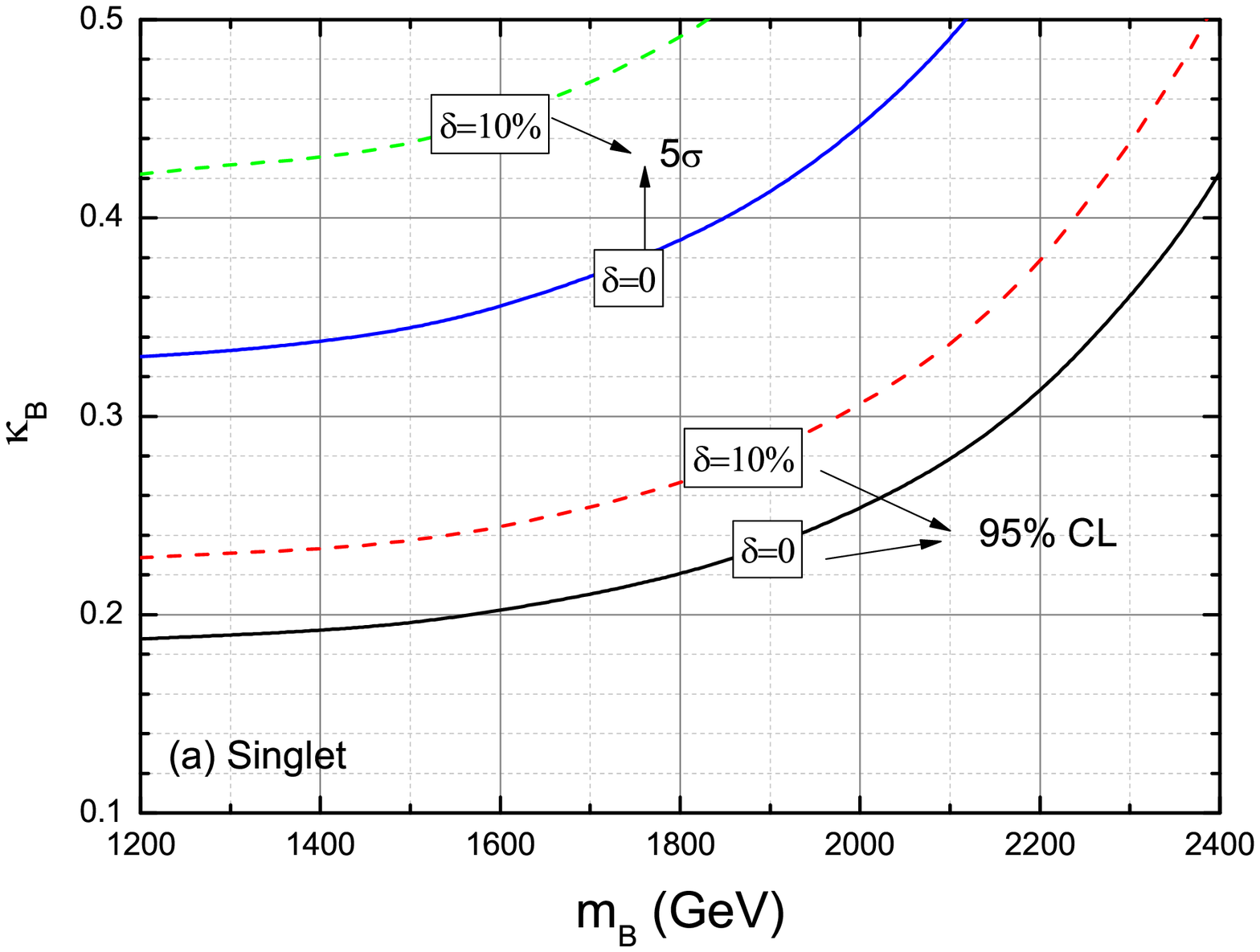}\epsfxsize=8cm \epsffile{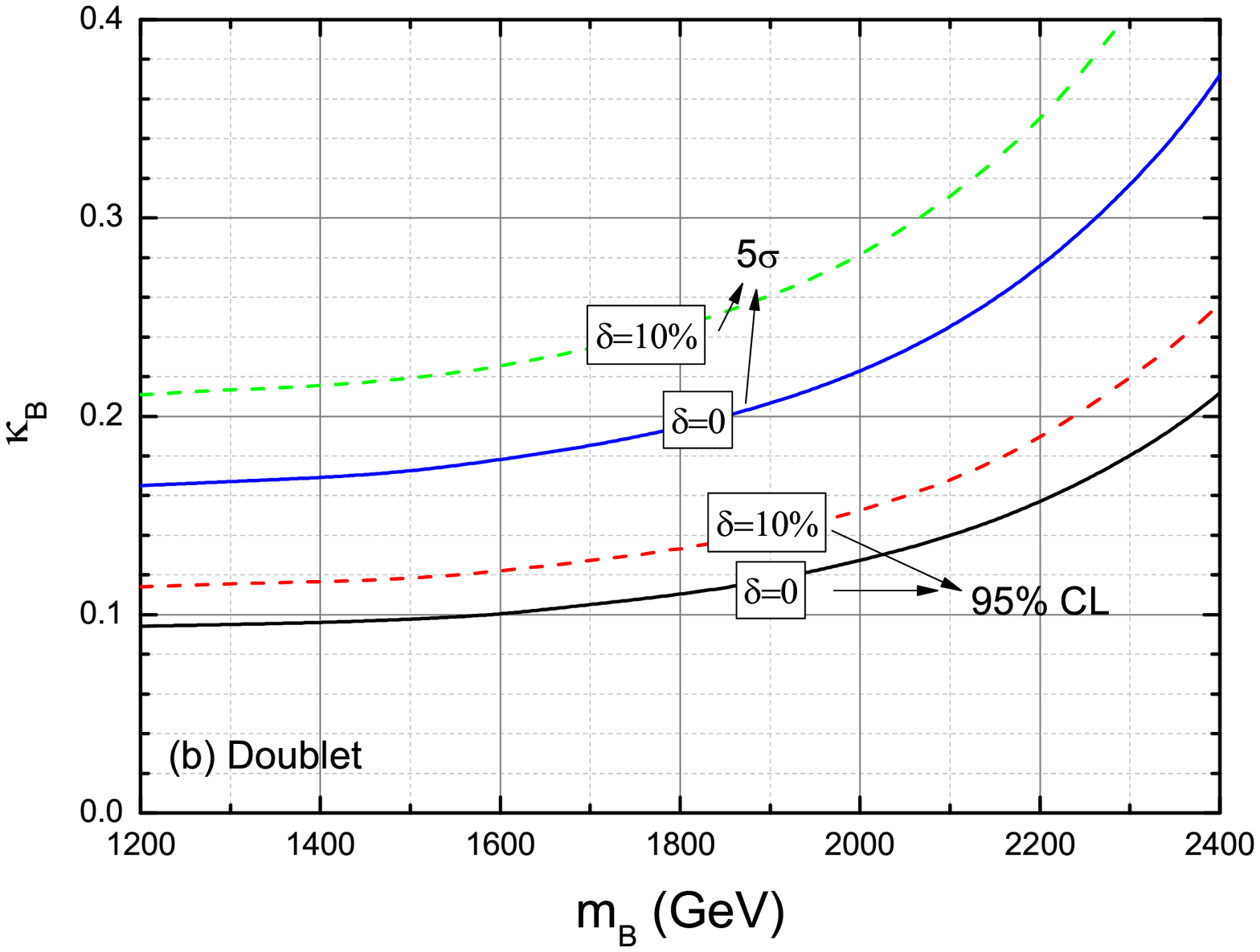}}
\caption{Combined  exclusion limit (at 95\% CL)  and discovery prospects (at $5\sigma$) contour plots for the signal in $\kappa_{B}-m_{B}$ planes at 3~TeV CLIC with an integral luminosity of 5 ab$^{-1}$ for the singlet~(left) and doublet~(right) cases. }
\label{fig7}
\end{center}
\end{figure}

Certainly, the sensitivities with some systematic errors will be weaker than those without any systematic error.
Next, we combine the significance with
$\mathcal{Z}_\text{comb}=\sqrt{\mathcal{Z}^{2}_{\ell\bar{\ell}}+\mathcal{Z}^{2}_{\nu\bar{\nu}} }$ by using the results from above two decay channels  with the aforementioned two systematic error cases of  $\delta=0$ and $\delta=10\%$.

In Fig.~\ref{fig7}, the combined 95\% CL exclusion limit and $5\sigma$ discovery prospect lines are drawn in $\kappa_{B}-m_{B}$ planes at the 3 TeV CLIC. One can see that, with a realistic 10\% systematic error, the sensitivities are slightly weaker than those without any systematic error.
For the singlet and doublet cases, the discovery region can, respectively,  reach  $\kappa_{B}\in [0.42, 0.52]$ with the VLQ-$B$ mass range
$m_B\in$ [1200 GeV, 1900 GeV], and $\kappa_{B}\in [0.21, 0.47]$ with the VLQ-$B$ mass range
$m_B\in$ [1200 GeV, 2400 GeV]. Otherwise, in the region of  $m_B\in$ [1200 GeV, 2400 GeV], the 95\% CL excluded region for the coupling parameter $\kappa_{B}$ is [0.23, 0.5] for the singlet case and [0.11, 0.25] for the doublet case, respectively,  at the 3 TeV CLIC with an integrated luminosity of 5 ab$^{-1}$.

%%% Fig.8 %%%%%%%%%%%%%%%%%%%%
\begin{figure}[htb]
\begin{center}
\vspace{-0.5cm}
\centerline{\epsfxsize=10cm \epsffile{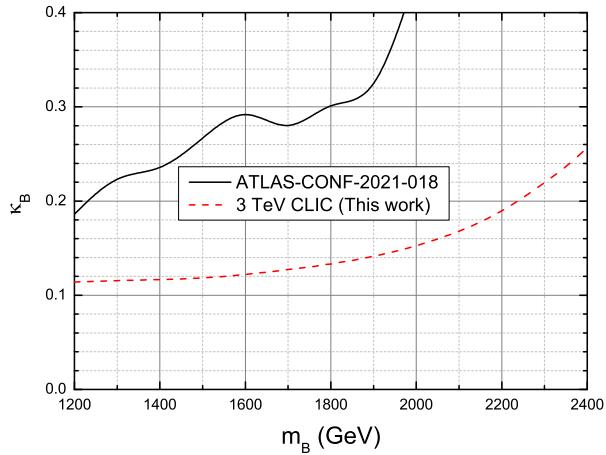}}
\caption{The contour of 95\% CL exclusion limit on the plane of $m_{B}$ versus $\kappa_{B}$ in the (B, Y) doublet scenarios with a realistic 10\% systematic error. The solid line indicates the  current expected limits from the LHC~\cite{ATLAS:2021gfv}. }
\label{fig8}
\end{center}
\end{figure}

The ATLAS Collaboration has  studied single production of a vector-like $B$ quark decaying into $bH$ with $H\to \gamma\gamma$ decay channel, assuming a generalized coupling $\kappa_B=0.5$ and doublet branching ratios of 50\% for $B\to hb$ and $B\to Zb$, $B$ quark with masses less than 1210 GeV are excluded at the 95\% CL~\cite{ATLAS:2018qxs}.
Very recently, the ATLAS Collaboration has presented the limit from the single
production of a VLQ-$B$ occurring as part of a $(B,Y)$ weak isospin doublet via $B\to bH(H\to b\bar{b})$ decay channel at 13 TeV LHC with 139 fb$^{-1}$ luminosity,
excluding $1.0 \rm ~TeV < m_B < 1.28 \rm ~TeV$ and $1.46 \rm ~TeV < m_B < 2.0 \rm ~TeV$ for $\kappa_B = 0.25$, and $1.0 \rm ~TeV < m_B < 2.0 \rm ~TeV$ for $\kappa_B = 0.3$~\cite{ATLAS:2021gfv}. In Fig.~\ref{fig8}, we give the current results at 13 TeV LHC and future reach at 3 TeV CLIC on the plane of $m_{B}$ versus $\kappa_{B}$ in
the $(B,Y)$ doublet scenarios.
 We can see that the future CLIC with $\sqrt{s}=3 \rm ~TeV$ and integrated luminosity of 5~ab$^{-1}$ could provide better sensitivity to detect the $BbZ$ couplings than the current  experimental results obtained from the current 13 TeV LHC.
  For high energy hadron colliders, the QCD backgrounds will be enhanced due to the pileup effect. By comparison, the properties of the VLQ-$B$ can be measured accurately once it is discovered due to the clean SM backgrounds at the future  leptonic colliders.

\section{Conclusion}
We have studied single production of VLQ-$B$ at the future 3 TeV CLIC via the process $e^{+}e^{-}\to B\bar{b}\to Zb\bar{b}$ in a model-independent way. We performed a full simulation for the signals and the relevant SM backgrounds based on two types of decay channels $Z\to \ell^{+}\ell^{-}$~$(\ell= e, \mu)$ and $Z\to \nu\bar{\nu}$.
The $5\sigma$ discovery prospects and 95\% CL exclusion limits in the parameter plane of the two variables $m_B$ and $\kappa_{B}$ were, respectively,  obtained at 3 TeV CLIC with an integral luminosity of 5~ab$^{-1}$.
Our numerical results show that, with the  systematic error case of  $\delta=10\%$, in the region of $m_B\in$ [1200 GeV, 2300 GeV], the discovery region can reach  $\kappa_{B}\in [0.42, 0.52]$ with the VLQ-$B$ mass range
$m_B\in$ [1200 GeV, 1900 GeV] for the singlet case, and $\kappa_{B}\in [0.21, 0.47]$ with the VLQ-$B$ mass range
$m_B\in$ [1200 GeV, 2400 GeV] for the doublet case, respectively.
Otherwise, in the region of  $m_B\in$ [1200 GeV, 2400 GeV], the excluded region for the coupling parameter $\kappa_{B}$ is [0.23, 0.5] for the singlet case and [0.11, 0.25] for the doublet case, respectively,  at the 3 TeV CLIC with an integrated luminosity of 5 ab$^{-1}$.

\begin{acknowledgments}
The work is supported by the Foundation of the Henan Science and Technology Research Project:202102210223; Key Research Projects in Universities of Henan:21A140031; Henan Colleges and Universities Youth Teacher Training Project:2020GGJS212.
\end{acknowledgments}


\begin{thebibliography}{99}
\bibitem{DeSimone:2012fs}
A.~De Simone, O.~Matsedonskyi, R.~Rattazzi and A.~Wulzer,
%``A First Top Partner Hunter's Guide,''
JHEP \textbf{04}, 004 (2013).
%doi:10.1007/JHEP04(2013)004
%[arXiv:1211.5663 [hep-ph]].

\bibitem{ArkaniHamed:2002qy}
  N.~Arkani-Hamed, A.~G.~Cohen, E.~Katz and A.~E.~Nelson,
  %``The Littlest Higgs,''
  JHEP {\bf 0207}, 034 (2002).
%  doi:10.1088/1126-6708/2002/07/034
%  [hep-ph/0206021].

\bibitem{Agashe:2006wa}
  K.~Agashe, G.~Perez and A.~Soni,
  %``Collider Signals of Top Quark Flavor Violation from a Warped Extra Dimension,''
  Phys.\ Rev.\ D {\bf 75}, 015002 (2007).
%  doi:10.1103/PhysRevD.75.015002
 % [hep-ph/0606293].

\bibitem{Agashe:2004rs}
K.~Agashe, R.~Contino and A.~Pomarol,
%``The Minimal composite Higgs model,''
Nucl. Phys. B \textbf{719}, 165-187 (2005).
%doi:10.1016/j.nuclphysb.2005.04.035
%[arXiv:hep-ph/0412089 [hep-ph]].



\bibitem{He:1999vp}
H.~J.~He, T.~M.~P.~Tait and C.~P.~Yuan,
%``New top flavor models with seesaw mechanism,''
Phys. Rev. D \textbf{62}, 011702 (2000).
%doi:10.1103/PhysRevD.62.011702
%[arXiv:hep-ph/9911266 [hep-ph]].
\bibitem{Wang:2013jwa}
X.~F.~Wang, C.~Du and H.~J.~He,
%``LHC Higgs Signatures from Topflavor Seesaw Mechanism,''
Phys. Lett. B \textbf{723}, 314-323 (2013).
%%doi:10.1016/j.physletb.2013.05.015
%%[arXiv:1304.2257 [hep-ph]].
\bibitem{He:2001fz}
H.~J.~He, C.~T.~Hill and T.~M.~P.~Tait,
%``Top Quark Seesaw, Vacuum Structure and Electroweak Precision Constraints,''
Phys. Rev. D \textbf{65}, 055006 (2002).
%doi:10.1103/PhysRevD.65.055006
%[arXiv:hep-ph/0108041 [hep-ph]].

\bibitem{He:2014ora}
H.~J.~He and Z.~Z.~Xianyu,
%``Extending Higgs Inflation with TeV Scale New Physics,''
JCAP \textbf{10}, 019 (2014).
%doi:10.1088/1475-7516/2014/10/019
%[arXiv:1405.7331 [hep-ph]].


\bibitem{Aguilar-Saavedra:2013qpa}
  J.~A.~Aguilar-Saavedra, R.~Benbrik, S.~Heinemeyer and M.~P\'erez-Victoria,
  %``Handbook of vectorlike quarks: Mixing and single production,''
  Phys.\ Rev.\ D {\bf 88}, 094010 (2013).
%  doi:10.1103/PhysRevD.88.094010
%  [arXiv:1306.0572 [hep-ph]].

\bibitem{Kribs:2007nz}
G.~D.~Kribs, T.~Plehn, M.~Spannowsky and T.~M.~P.~Tait,
%``Four generations and Higgs physics,''
Phys. Rev. D \textbf{76}, 075016 (2007).
%doi:10.1103/PhysRevD.76.075016
%[arXiv:0706.3718 [hep-ph]].


\bibitem{Banerjee:2013hxa}
S.~Banerjee, M.~Frank and S.~K.~Rai,
%``Higgs data confronts Sequential Fourth Generation Fermions in the Higgs Triplet Model,''
Phys. Rev. D \textbf{89}, no.7, 075005 (2014).
%doi:10.1103/PhysRevD.89.075005
%[arXiv:1312.4249 [hep-ph]].
%%%%%%%%%%%%for example%%%%%%%%%%%%%%%%%%%%%%%%


\bibitem{Buchkremer:2013bha}
  M.~Buchkremer, G.~Cacciapaglia, A.~Deandrea and L.~Panizzi,
  %``Model Independent Framework for Searches of Top Partners,''
  Nucl.\ Phys.\ B {\bf 876}, 376 (2013).
%  doi:10.1016/j.nuclphysb.2013.08.010
 % [arXiv:1305.4172 [hep-ph]].

\bibitem{Barducci:2017xtw}
D.~Barducci and L.~Panizzi,
%``Vector-like quarks coupling discrimination at the LHC and future hadron colliders,''
JHEP \textbf{12}, 057 (2017).
%doi:10.1007/JHEP12(2017)057
%[arXiv:1710.02325 [hep-ph]].

\bibitem{Cacciapaglia:2018qep}
G.~Cacciapaglia, A.~Carvalho, A.~Deandrea, T.~Flacke, B.~Fuks, D.~Majumder, L.~Panizzi and H.~S.~Shao,
%``Next-to-leading-order predictions for single vector-like quark production at the LHC,''
Phys. Lett. B \textbf{793}, 206-211 (2019).
%doi:10.1016/j.physletb.2019.04.056
%[arXiv:1811.05055 [hep-ph]].
\bibitem{Fuks:2016ftf}
B.~Fuks and H.~S.~Shao,
%``QCD next-to-leading-order predictions matched to parton showers for vector-like quark models,''
Eur. Phys. J. C \textbf{77}, no.2, 135 (2017).
%doi:10.1140/epjc/s10052-017-4686-z
%[arXiv:1610.04622 [hep-ph]].



\bibitem{Yang:2014usa}
S.~Yang, J.~Jiang, Q.~S.~Yan and X.~Zhao,
%``Hadronic b' search at the LHC with top and W taggers,''
JHEP \textbf{09}, 035 (2014).
%doi:10.1007/JHEP09(2014)035
%[arXiv:1405.2514 [hep-ph]].

\bibitem{Liu:2018hum}
D.~Liu, L.~T.~Wang and K.~P.~Xie,
%``Prospects of searching for composite resonances at the LHC and beyond,''
JHEP \textbf{01}, 157 (2019).
%doi:10.1007/JHEP01(2019)157
%[arXiv:1810.08954 [hep-ph]].

\bibitem{Cacciapaglia:2018lld}
G.~Cacciapaglia, A.~Deandrea, N.~Gaur, D.~Harada, Y.~Okada and L.~Panizzi,
%``The LHC potential of Vector-like quark doublets,''
JHEP \textbf{11}, 055 (2018).
%doi:10.1007/JHEP11(2018)055
%[arXiv:1806.01024 [hep-ph]].

\bibitem{Aguilar-Saavedra:2019ghg}
J.~A.~Aguilar-Saavedra, J.~Alonso-Gonz\'alez, L.~Merlo and J.~M.~No,
%``Exotic vectorlike quark phenomenology in the minimal linear \ensuremath{\sigma} model,''
Phys. Rev. D \textbf{101}, no.3, 035015 (2020).
%doi:10.1103/PhysRevD.101.035015
%[arXiv:1911.10202 [hep-ph]].

\bibitem{Wang:2020ips}
D.~Wang, L.~Wu and M.~Zhang,
%``Hunting for top partner with a new signature at the LHC,''
Phys. Rev. D \textbf{103}, no.11, 115017 (2021).
%doi:10.1103/PhysRevD.103.115017
%[arXiv:2007.09722 [hep-ph]].

%%%%%%%%%%%%%%%%%%%%%%%%%%%%%%%%%%%%%%%%%%%%%%%%%%%%%


\bibitem{Zhang:2017nsn}
Y.~J.~Zhang, L.~Han and Y.~B.~Liu,
%``Single production of the top partner in the T \textrightarrow{} tZ channel at the LHeC,''
Phys. Lett. B \textbf{768}, 241-247 (2017).
%doi:10.1016/j.physletb.2017.02.051

\bibitem{Han:2017cvu}
L.~Han, Y.~J.~Zhang and Y.~B.~Liu,
%``Single vector-like $T$-quark search via the $T \to Wb$ decay channel at the LHeC,''
Phys. Lett. B \textbf{771}, 106-112 (2017).
%doi:10.1016/j.physletb.2017.05.036

\bibitem{Liu:2017rjw}
Y.~B.~Liu,
%``Search for single production of vector-like top partners at the Large Hadron Electron Collider,''
Nucl. Phys. B \textbf{923}, 312-323 (2017).
%doi:10.1016/j.nuclphysb.2017.08.006
%[arXiv:1704.02059 [hep-ph]].

\bibitem{Liu:2017sdg}
Y.~B.~Liu and Y.~Q.~Li,
%``Search for single production of the vector-like top partner at the 14 TeV LHC,''
Eur. Phys. J. C \textbf{77}, no.10, 654 (2017).
%doi:10.1140/epjc/s10052-017-5228-4
%[arXiv:1709.06427 [hep-ph]].


\bibitem{Liu:2019jgp}
Y.~B.~Liu and S.~Moretti,
%``Search for single production of a top quark partner via the $T\to th$ and $h\to WW^{\ast}$ channels at the LHC,''
Phys. Rev. D \textbf{100}, no.1, 015025 (2019).
%doi:10.1103/PhysRevD.100.015025
%[arXiv:1902.03022 [hep-ph]].

\bibitem{Tian:2021oey}
X.~Y.~Tian, L.~F.~Du and Y.~B.~Liu,
%``Single production of vector-like top partners in trilepton channel at future 100 TeV Hadron colliders,''
Nucl. Phys. B \textbf{965}, 115358 (2021).
%doi:10.1016/j.nuclphysb.2021.115358





\bibitem{Moretti:2016gkr}
S.~Moretti, D.~O'Brien, L.~Panizzi and H.~Prager,
%``Production of extra quarks at the Large Hadron Collider beyond the Narrow Width Approximation,''
Phys. Rev. D \textbf{96}, no.7, 075035 (2017).
%doi:10.1103/PhysRevD.96.075035
%[arXiv:1603.09237 [hep-ph]].

\bibitem{Moretti:2017qby}
S.~Moretti, D.~O'Brien, L.~Panizzi and H.~Prager,
%``Production of extra quarks decaying to Dark Matter beyond the Narrow Width Approximation at the LHC,''
Phys. Rev. D \textbf{96}, no.3, 035033 (2017).
%doi:10.1103/PhysRevD.96.035033
%[arXiv:1705.07675 [hep-ph]].

\bibitem{Carvalho:2018jkq}
A.~Carvalho, S.~Moretti, D.~O'Brien, L.~Panizzi and H.~Prager,
%``Single production of vectorlike quarks with large width at the Large Hadron Collider,''
Phys. Rev. D \textbf{98}, no.1, 015029 (2018).
%doi:10.1103/PhysRevD.98.015029
%[arXiv:1805.06402 [hep-ph]].



\bibitem{Roy:2020fqf}
A.~Roy, N.~Nikiforou, N.~Castro and T.~Andeen,
%``Novel interpretation strategy for searches of singly produced vectorlike quarks at the LHC,''
Phys. Rev. D \textbf{101}, no.11, 115027 (2020).
%doi:10.1103/PhysRevD.101.115027
%[arXiv:2003.00640 [hep-ph]].

 \bibitem{Buckley:2020wzk}
A.~Buckley, J.~M.~Butterworth, L.~Corpe, D.~Huang and P.~Sun,
%``New sensitivity of current LHC measurements to vector-like quarks,''
SciPost Phys. \textbf{9}, no.5, 069 (2020).
%doi:10.21468/SciPostPhys.9.5.069
%[arXiv:2006.07172 [hep-ph]].



%\cite{Deandrea:2021vje}
\bibitem{Deandrea:2021vje}
A.~Deandrea, T.~Flacke, B.~Fuks, L.~Panizzi and H.~S.~Shao,
%``Single production of vector-like quarks: the effects of large width, interference and NLO corrections,''
JHEP \textbf{08}, 107 (2021).
%doi:10.1007/JHEP08(2021)107
%[arXiv:2105.08745 [hep-ph]].

\bibitem{King:2021iah}
S.~J.~D.~King, S.~F.~King, S.~Moretti and S.~J.~Rowley,
%``Discovering the Origin of Yukawa Couplings at the LHC with a Singlet Higgs and Vector-like Quarks,''
JHEP \textbf{21}, 144 (2020).
%doi:10.1007/JHEP05(2021)144
%[arXiv:2102.06091 [hep-ph]].

\bibitem{Atre:2011ae}
  A.~Atre, G.~Azuelos, M.~Carena, T.~Han, E.~Ozcan, J.~Santiago and G.~Unel,
  %``Model-Independent Searches for New Quarks at the LHC,''
  JHEP {\bf 1108}, 080 (2011).
%  doi:10.1007/JHEP08(2011)080
%  [arXiv:1102.1987 [hep-ph]].
%%%%%%%%%%%%%%%%ATLAS and CMS search%%%%%%%%%%%%%%%%%

  \bibitem{Aaboud:2018wxv}
  M.~Aaboud {\it et al.} [ATLAS Collaboration],
  %``Search for pair production of heavy vector-like quarks decaying into hadronic final states in $pp$ collisions at $\sqrt{s} = 13$ TeV with the ATLAS detector,''
  Phys.\ Rev.\ D {\bf 98}, 092005 (2018).
%  doi:10.1103/PhysRevD.98.092005
%  [arXiv:1808.01771 [hep-ex]].


\bibitem{Aaboud:2018xpj}
  M.~Aaboud {\it et al.} [ATLAS Collaboration],
  %``Search for new phenomena in events with same-charge leptons and $b$-jets in $pp$ collisions at $\sqrt{s}= 13$ TeV with the ATLAS detector,''
  JHEP {\bf 1812}, 039 (2018).
%  doi:10.1007/JHEP12(2018)039
%  [arXiv:1807.11883 [hep-ex]].

\bibitem{Aaboud:2018uek}
  M.~Aaboud {\it et al.} [ATLAS Collaboration],
  %``Search for pair production of heavy vector-like quarks decaying into high-$p_T$ $W$ bosons and top quarks in the lepton-plus-jets final state in $pp$ collisions at $\sqrt{s}=13$ TeV with the ATLAS detector,''
  JHEP {\bf 1808}, 048 (2018).
%   doi:10.1007/JHEP08(2018)048
%   [arXiv:1806.01762 [hep-ex]].


 \bibitem{Aaboud:2018ifs}
  M.~Aaboud {\it et al.} [ATLAS Collaboration],
  %``Search for single production of vector-like quarks decaying into $Wb$ in $pp$ collisions at $\sqrt{s} = 13$ TeV with the ATLAS detector,''
  JHEP {\bf 1905}, 164 (2019).
%   doi:10.1007/JHEP05(2019)164
%   [arXiv:1812.07343 [hep-ex]].


\bibitem{Sirunyan:2018qau}
  A.~M.~Sirunyan {\it et al.} [CMS Collaboration], ~Eur.\ Phys.\ J.\ C {\bf 79}, 364 (2019).
  %%1812.09768


\bibitem{Sirunyan:2019sza}
  A.~M.~Sirunyan {\it et al.} [CMS Collaboration],
  %``Search for pair production of vectorlike quarks in the fully hadronic final state,''
  Phys.\ Rev.\ D {\bf 100}, 072001 (2019).
%  doi:10.1103/PhysRevD.100.072001
 % [arXiv:1906.11903 [hep-ex]].

  \bibitem{Sirunyan:2018omb}
  A.~M.~Sirunyan {\it et al.} [CMS Collaboration], JHEP \textbf{08}, 177 (2018).
  %%1805.04758
\bibitem{Sirunyan:2020qvb}
  A.~M.~Sirunyan {\it et al.} [CMS Collaboration],
  %``Search for pair production of vectorlike quarks in the fully hadronic final state,''
  Phys.\ Rev.\ D {\bf 102},112004 (2020).
  %%%2008.09835 [hep-ex]
  \bibitem{Aaboud:2018pii}
  M.~Aaboud {\it et al.} [ATLAS Collaboration],
  %``Combination of the searches for pair-produced vector-like partners of the third-generation quarks at $\sqrt{s} =$ 13 TeV with the ATLAS detector,''
  Phys.\ Rev.\ Lett.\  {\bf 121}, 211801 (2018).
%  doi:10.1103/PhysRevLett.121.211801
%  [arXiv:1808.02343 [hep-ex]].

%%%%%%%%%%%%%%%%%%%%%%%%%%%%%%%%%%%%%%%%%%%%%%%%
\bibitem{Nutter:2012an}
J.~Nutter, R.~Schwienhorst, D.~G.~E.~Walker and J.~H.~Yu,
%``Single Top Production as a Probe of B-prime Quarks,''
Phys. Rev. D \textbf{86}, 094006 (2012).
%%doi:10.1103/PhysRevD.86.094006
%%[arXiv:1207.5179 [hep-ph]].
\bibitem{Gong:2019zws}
X.~Gong, C.~X.~Yue and Y.~C.~Guo,
Phys.\ Lett.\ B {\bf 793}, 175 (2019).
\bibitem{Gong:2020ouh}
X.~Gong, C.~X.~Yue, H.~M. Yu and D.~Li,
Eur.\ Phys.\ J. C {\bf 80}, 876 (2020).
%%%%%%%%%%%%%%%%%%%%%%%%%%%%%%%%%%%%%%%%%%%%%%%%%%%%%%%%%%55
\bibitem{CLIC1}
H.~Abramowicz \textit{et al.} [CLIC Detector and Physics Study],
%``Physics at the CLIC e+e- Linear Collider -- Input to the Snowmass process 2013,''
arXiv:1307.5288 [hep-ex].
\bibitem{CLIC2}
J.~de Blas, R.~Franceschini, F.~Riva, P.~Roloff, U.~Schnoor, M.~Spannowsky, J.~D.~Wells, A.~Wulzer, J.~Zupan and S.~Alipour-Fard, \textit{et al.}
%``The CLIC Potential for New Physics,''
%%doi:10.23731/CYRM-2018-003
arXiv:1812.02093 [hep-ph].
\bibitem{CLIC3}
R.~Franceschini,
%``Beyond the Standard Model physics at CLIC,''
Int. J. Mod. Phys. A \textbf{35}, 2041015 (2020).
%%doi:10.1142/S0217751X20410158
%%[arXiv:1902.10125 [hep-ph]].
\bibitem{Dannheim:2013ypa}
D.~Dannheim, P.~Lebrun, L.~Linssen, D.~Schulte and S.~Stapnes,
%``CLIC $e^+ e^-$ Linear Collider Studies - Input to the Snowmass process 2013,''
arXiv:1305.5766 [physics.acc-ph].


\bibitem{Kitano:2002ss}
R.~Kitano, T.~Moroi and S.~f.~Su,
%``Top squark study at a future e+ e- linear collider,''
JHEP \textbf{12}, 011 (2002).
%doi:10.1088/1126-6708/2002/12/011
%[arXiv:hep-ph/0208149 [hep-ph]].
\bibitem{Kong:2007uu}
K.~Kong and S.~C.~Park,
%``Phenomenology of Top partners at the ILC,''
JHEP \textbf{08}, 038 (2007).
%doi:10.1088/1126-6708/2007/08/038
%[arXiv:hep-ph/0703057 [hep-ph]].
\bibitem{Senol:2011nm}
A.~Senol, A.~T.~Tasci and F.~Ustabas,
%``Anomalous single production of fourth generation $t'$ quarks at ILC and CLIC,''
Nucl. Phys. B \textbf{851}, 289-297 (2011).
%doi:10.1016/j.nuclphysb.2011.05.022
%[arXiv:1104.5316 [hep-ph]].
\bibitem{Harigaya:2011yg}
K.~Harigaya, S.~Matsumoto, M.~M.~Nojiri and K.~Tobioka,
%``Testing Little Higgs Mechanism at Future Colliders,''
JHEP \textbf{01}, 135 (2012).
%doi:10.1007/JHEP01(2012)135
%[arXiv:1109.4847 [hep-ph]].
\bibitem{Guo:2014piv}
M.~A.B., L.~Guo, W.~Liu, W.~G.~Ma, R.~Y.~Zhang and W.~J.~Zhang,
%``Precision calculations for the $T$-odd quark pair production at the CLIC $e^+e^-$ linear collider,''
Commun. Theor. Phys. \textbf{62}, no.6, 824-832 (2014).
%[arXiv:1406.5256 [hep-ph]].
\bibitem{Liu:2014pts}
Y.~B.~Liu and Z.~J.~Xiao,
%``The production and decay of the top partner T in the left\textendash{}right twin Higgs model at the ILC and CLIC,''
Nucl. Phys. B \textbf{892}, 63-82 (2015).
%doi:10.1016/j.nuclphysb.2014.12.027
%[arXiv:1409.6050 [hep-ph]].

\bibitem{Qin:2021cxl}
X.~Qin and J.~F.~Shen,
%``Search for single production of vector-like $B$ quark decaying to a Higgs boson and bottom quark at the CLIC,''
Nucl. Phys. B \textbf{966}, 115388 (2021).
%%doi:10.1016/j.nuclphysb.2021.115388
\bibitem{Han:2021kcr}
L.~Han and J.~F.~Shen,
%``Search for single production of vector-like $B$ quark decaying to $bZ$ at future linear colliders,''
Eur. Phys. J. C \textbf{81}, no.5, 463 (2021).
%doi:10.1140/epjc/s10052-021-09245-y
\bibitem{Atre:2008iu}
A.~Atre, M.~Carena, T.~Han and J.~Santiago,
%``Heavy Quarks Above the Top at the Tevatron,''
Phys. Rev. D \textbf{79}, 054018 (2009).
%doi:10.1103/PhysRevD.79.054018
%[arXiv:0806.3966 [hep-ph]].

%%%%%%%%%%%%%%%%%%%%%%%%%%%%%%%%%%%%%%%%%%%%%%%%%%%%%%%%%%%%%%%%%%333333333333333333333333333333%%%%%%%%%%%%%%%%%%%%%%%%%%%%%%%
\bibitem{ET-hjh}
 For a comprehensive review,
 H.~J.~He, Y.~P.~Kuang and C.~P.~Yuan,
%``Global analysis for probing electroweak symmetry breaking mechanism at high-energy colliders,''
[arXiv:hep-ph/9704276].
\bibitem{He:1992nga}
H.~J.~He, Y.~P.~Kuang and X.~y.~Li,
%``On the precise formulation of equivalence theorem,''
Phys. Rev. Lett. \textbf{69}, 2619-2622 (1992).
%doi:10.1103/PhysRevLett.69.2619
%163 citations counted in INSPIRE as of 02 Feb 2021
\bibitem{He:1993yd}
H.~J.~He, Y.~P.~Kuang and X.~y.~Li,
%``Further investigation on the precise formulation of the equivalence theorem,''
Phys. Rev. D \textbf{49}, 4842-4872 (1994).
%doi:10.1103/PhysRevD.49.4842
%100 citations counted in INSPIRE as of 13 Feb 2021
\bibitem{He:1994br}
H.~J.~He, Y.~P.~Kuang and C.~P.~Yuan,
%``Equivalence theorem and probing the electroweak symmetry breaking sector,''
Phys. Rev. D \textbf{51}, 6463-6473 (1995).
%doi:10.1103/PhysRevD.51.6463
%[arXiv:hep-ph/9410400 [hep-ph]].
%% [arXiv:hep-ph/9410400];
\bibitem{He:1996rb}
H.~J.~He, Y.~P.~Kuang and C.~P.~Yuan,
%``Estimating the sensitivity of LHC to electroweak symmetry breaking: Longitudinal / Goldstone boson equivalence as a criterion,''
Phys. Rev. D \textbf{55}, 3038-3067 (1997).
%doi:10.1103/PhysRevD.55.3038
%[arXiv:hep-ph/9611316 [hep-ph]].
\bibitem{He:1996cm}
H.~J.~He and W.~B.~Kilgore,
%``The Equivalence theorem and its radiative correction - free formulation for all R(xi) gauges,''
Phys. Rev. D \textbf{55}, 1515-1532 (1997).
%%doi:10.1103/PhysRevD.55.1515
%%[arXiv:hep-ph/9609326 [hep-ph]].




%%%%%%%%%%%%%%%%%%%%%%%%%%%%%%%%%%%%%%%%%%%%%%%%%%%%%%%%%%%%%%%%%%333333333333333333333333333333%%%%%%%%%%%%%%%%%%%%%%%%%%%%%%%
\bibitem{mg5}
J. Alwall, R. Frederix, S. Frixione, V. Hirschi, F. Maltoni, O. Mattelaer, H.-S. Shao, T. Stelzer, P. Torrielli and M. Zaro, JHEP {\bf 1407}, 079 (2014).
%``The automated computation of tree-level and next-to-leading order differential cross sections, and their matching to parton shower simulations,''
%%  doi:10.1007/JHEP07(2014)079
%%  [arXiv:1405.0301 [hep-ph]].

\bibitem{pdg}
M. Tanabashi {\it et al.}, [Particle Data Group],  Phys.\ Rev.\ D {\bf 98}, 030001 (2018).
%%{\it The Review of Particle Physics},


\bibitem{pythia8}
T.~Sj\"ostrand, S. Ask and J. R. Christiansen {\it et al.},  Comput. Phys. Commun. {\bf 191}, 159 (2015).
  %``An Introduction to PYTHIA 8.2,''
%%  doi:10.1016/j.cpc.2015.01.024
%%  [arXiv:1410.3012 [hep-ph]].


\bibitem{deFavereau:2013fsa}
J.~de Favereau \textit{et al.} [DELPHES 3],
%``DELPHES 3, A modular framework for fast simulation of a generic collider experiment,''
JHEP \textbf{02}, 057 (2014).
%doi:10.1007/JHEP02(2014)057
%[arXiv:1307.6346 [hep-ex]].
%1743 citations counted in INSPIRE as of 31 Dec 2020
\bibitem{Leogrande:2019qbe}
E.~Leogrande, P.~Roloff, U.~Schnoor and M.~Weber,
%``A DELPHES card for the CLIC detector,''
arXiv:1909.12728 [hep-ex].

\bibitem{Boronat:2014hva}
M.~Boronat, J.~Fuster, I.~Garcia, E.~Ros and M.~Vos,
%``A robust jet reconstruction algorithm for high-energy lepton colliders,''
Phys. Lett. B \textbf{750}, 95-99 (2015).
%doi:10.1016/j.physletb.2015.08.055
%[arXiv:1404.4294 [hep-ex]].
\bibitem{Boronat:2016tgd}
M.~Boronat, J.~Fuster, I.~Garcia, P.~Roloff, R.~Simoniello and M.~Vos,
%``Jet reconstruction at high-energy electron\textendash{}positron colliders,''
Eur.\ Phys.\ J.\ C \textbf{78}, 144 (2018).
%doi:10.1140/epjc/s10052-018-5594-6
%[arXiv:1607.05039 [hep-ex]].

\bibitem{ma5}
E.~Conte, B.~Fuks and G.~Serret,
%``MadAnalysis 5, A User-Friendly Framework for Collider Phenomenology,''
Comput. Phys. Commun. \textbf{184}, 222-256 (2013).
%doi:10.1016/j.cpc.2012.09.009
%[arXiv:1206.1599 [hep-ph]].
%363 citations counted in INSPIRE as of 31 Dec 2020

\bibitem{Cowan:2010js}
  G.~Cowan, K.~Cranmer, E.~Gross and O.~Vitells,
  %``Asymptotic formulae for likelihood-based tests of new physics,''
  Eur.\ Phys.\ J.\ C {\bf 71}, 1554 (2011),
  Erratum: [Eur.\ Phys.\ J.\ C {\bf 73} (2013) 2501]
%%  doi:10.1140/epjc/s10052-011-1554-0, 10.1140/epjc/s10052-013-2501-z
%%  [arXiv:1007.1727 [physics.data-an]].
\bibitem{ATLAS:2018qxs}
 [ATLAS],
%``Search for single production of a vector-like $B$ quark decaying into a bottom quark and a Higgs boson which decays into a pair of photons,''
ATLAS-CONF-2018-024.
%13 citations counted in INSPIRE as of 08 Nov 2021
\bibitem{ATLAS:2021gfv}
 [ATLAS],
%``Search for single vector-like $B$ quark production and decay via $B\rightarrow bH(b\bar{b})$ in pp collisions at $\sqrt{s} = 13\text{ TeV}$ with the ATLAS detector,''
ATLAS-CONF-2021-018.
%1 citations counted in INSPIRE as of 08 Nov 2021
\end{thebibliography}
\end{document}